\documentclass[journal]{IEEEtran}
\usepackage{graphicx}
\usepackage{subfigure}
\usepackage{float}
\pdfoutput=1
\usepackage{amsmath}
\usepackage{epstopdf}
\usepackage{cases}
\usepackage{color}
\usepackage{colortbl}
\usepackage{stfloats}
\usepackage{ulem} 
\usepackage{algorithm}
\usepackage{algpseudocode}
\usepackage{amsmath}
\usepackage{amssymb}
\usepackage{booktabs}
\usepackage{setspace}
\usepackage{array}
\makeatletter
\renewcommand{\maketag@@@}[1]{\hbox{\m@th\normalsize\normalfont#1}}%
\makeatother
\usepackage{stfloats}
\usepackage{cite}
\usepackage{makecell}
\usepackage{multirow}
\usepackage{hyperref}

\ifCLASSINFOpdf
\else
\fi

\usepackage{caption}
\usepackage{mathtools}

\begin{document}

\title{Carrier Aggregation Enabled Integrated Sensing and Communication Signal Design and Processing}
\author{Zhiqing~Wei,
	Haotian~Liu, Xinyi~Yang, Wangjun~Jiang,\\
	Huici~Wu,
	Xingwang~Li,
	Zhiyong~Feng}

\maketitle

\begin{abstract}
The future mobile 
communication systems will support intelligent applications such as 
Internet of Vehicles (IoV) and Extended Reality (XR).
Integrated Sensing and Communication (ISAC) is regarded as one of the
key technologies satisfying the 
high data rate communication and highly accurate sensing for these intelligent applications
in future mobile communication systems. 
With the explosive growth of wireless devices and services, 
the shortage of spectrum resources leads to the fragmentation of available frequency 
bands for ISAC systems, 
which degrades sensing performance. 
Facing the above challenges, 
this paper proposes a Carrier Aggregation (CA)-based ISAC signal aggregating high and low-frequency bands 
to improve the sensing performance, where the CA-based ISAC signal can use four different aggregated pilot structures for sensing. 
Then, an ISAC signal processing algorithm with Compressed Sensing (CS)
is proposed and the Fast Iterative Shrinkage-Thresholding Algorithm (FISTA) is used to solve the reconfiguration convex optimization problem.
Finally, the Cram\'{e}r-Rao Lower Bounds (CRLBs) are derived
for the CA-based ISAC signal.
Simulation results show that CA
efficiently improves the accuracy of range and velocity estimation.
\end{abstract}


\IEEEpeerreviewmaketitle

\section{Introduction}

\begin{table*}
	\caption{Abbreviations and Notations}
	\begin{center}
        \renewcommand{\arraystretch}{1.3}
		\begin{tabular}{m{0.09\textwidth}<{\centering} m{0.37\textwidth}<{} m{0.09\textwidth}<{\centering} m{0.37\textwidth}<{}}
			\hline
			\hline			
			{Abbreviation} & {Description}&{Abbreviation} & {Description} \\
			\hline 
			5G-A&5th-Generation-Advanced &
			6G&6th-Generation  \\
                2D&Two-dimensional &
                3GPP& 3rd Generation Partnership Project \\
			AWGN& Additive White Gaussian Noise &
			ADC & Analog-to-digital converter \\
			BS& Base station&
			CRLB&Cram\'{e}r-Rao Lower Bound \\
			CCs&Component Carriers &
			CSI& Channel state information \\
			CS&Compressed Sensing &
			CP& Cyclic Prefix \\
			CA&Carrier aggregation  &
			DFT&Discrete Fourier Transform\\
			DAC & Digital-to-analog converter &
			FFT&  Fast Fourier Transform \\
                FISTA& Fast iterative shrinkage-thresholding algorithm &
			IDFT&Inverse Discrete Fourier Transform\\
			IFFT& Inverse Fast Fourier Transform &
			ISAC&Integrated sensing and communication\\
			  IoV& Internet of Vehicle&
			  IM& Index-modulation\\
			ISI & Inter-symbol interference &
                LO& Local oscillator\\
                mmWave&Millimeter wave&
                MAC&Media Access Control\\
                OMP & Orthogonal Matching Pursuit&
                OFDM & Orthogonal Frequency Division Multiplexing \\
                P/S & Parallel-to-serial conversion &
			RX& Receiver \\
			RCS& Radar Cross Section&
                RIP& Restricted Isometry Property \\
                RMSE & Root Mean Square Error &
                RCRLB & Root Cram\'{e}r-Rao Lower Bound \\
			S/P & Serial-to-parallel conversion &
                SNR&Signal-to-Noise Ratio\\
                TX& Transmitter&
                XR&Extended Reality\\ 
			\hline
			\hline
			{Notation} & {Description}&{Notation} & {Description} \\
			
			\hline
			${N}$ & Total number of subcarriers &
			${M}$ & Total number of OFDM symbols\\
			${k, K}$ & Interval of comb pilot  &
			${q, Q}$ & Interval of block pilot\\
			${T_1}$ &Total length of symbols in low-frequency&
			${T_2}$ &Total length of symbols in high-frequency\\
			${\Delta f_1}$ & Subcarrier spacing in low-frequency &
			${\Delta f_2}$ & Subcarrier spacing in high-frequency\\
			${f_{c1}}$ &Carrier frequency in low-frequency&
			${f_{c2}}$ & Carrier frequency in high-frequency\\
			${T_{CP}}$ & Length of CP &
			$\mathbf{D}_{cf1}$ & Channel information matrix in low-frequency \\
			$\mathbf{D}_{cf2}$ & Channel information matrix in high-frequency &
			$R, v_0$ & Range and velocity of target\\
			$\textbf{Q},\textbf{P}$ &Selection matrix &
			$ (\cdot)^\ast$& Complex conjugate of the complex number\\
			$(\cdot)^\mathrm{H}, (\cdot)^\mathrm{T}$ & Conjugate transpose, tanspose &
			$\mathrm{vec}(\cdot)$ & Stacking matrix in columns \\
			$\circ $ & Hadamard product &
			$ \otimes $ & Kronecker product\\
			\hline
			\hline
		\end{tabular}
	\end{center}
	\label{table_0}
\end{table*}

The future 5th-Generation-Advanced (5G-A) and 6th-Generation (6G) mobile 
communication systems will support intelligent applications such as 
Internet of Vehicles (IoV) and Extended Reality (XR) \cite{wei2023integrated}. 
These applications require both high data rate communication 
and highly accurate sensing. 
Designing a system with both communication 
and sensing functions is a promising solution to meet the above requirements \cite{kumari2017ieee}.
With the rapid development of mobile communication systems, 
the frequency 
bands of communication systems are rising, which are getting close to the 
frequency bands of radar systems. 
Moreover, the signal processing methods 
and hardware structures of radar and communication systems are similar \cite{feng2020joint}.
Hence, it is feasible to realize the integrated design of radar sensing and communication. 
The remarkable results of recent researches \cite{feng2020joint,wang2021symbiotic,li2022novel,yuan2021integrated} 
show that Integrated Sensing and Communication (ISAC) has the advantages of 
improving spectrum utilization and reducing device size, 
which is one of the promising key technologies in 5G-A and 6G.

Since the frequency bands of mobile communication systems are congested, 
the available spectrum resources for ISAC systems are fragmented, 
which will degrade the sensing performance. 
On one hand, there are deviations 
in sensing information such as Doppler frequency shifts carried by the echo signals 
in different frequency bands, 
making it difficult to directly apply multi-signal accumulation to improve the 
Signal-to-Noise Ratio (SNR) of the echo signals. 
On the other hand, when traditional radar sensing algorithms 
are applied to extract the sensing information from the echo signals on 
the fragmented spectrum resources, 
the fragmented spectrum will increase the level of sidelobe and 
decrease the sensing accuracy \cite{1021843429.nh}.

Hence, the efficient aggregation of the fragmented spectrum resources 
in radar sensing is a great challenge of ISAC signal processing on the 
congested frequency bands. 
In the field of communication, Carrier Aggregation (CA) 
techniques enable the aggregation of fragmented spectrum resources. 
CA is a spectrum expansion technique adopted by 3rd Generation Partnership Project (3GPP) 
to alleviate the 
shortage of contiguous resources. 
CA could be implemented in physical layer 
or Media Access Control (MAC) layer, which is divided into two categories, 
namely physical layer CA and MAC layer CA \cite{CAexplained}. 
In physical layer CA, the 
selected Component Carriers (CCs) share a common transmission module, 
where 
the same modulation and coding techniques are applied. 
In MAC layer CA, the 
selected CCs use different transmission modules, where different modulation 
and coding techniques could be applied depending on the 
current Channel State Information (CSI) \cite{ratasuk2010carrier}. 
In terms of communication performance, 
according to the equation of channel capacity \cite{goldsmith2005wireless}, 
CA significantly improves the channel capacity by increasing the bandwidth.
According to \cite{thoma2021joint,wild2021joint}, in the field of ISAC signal design and processing, it is expected to introduce the idea of CA to aggregate fragmented spectrum resources to improve the resolution and accuracy of radar sensing.

\begin{figure}[!h]
	\centering
	\includegraphics[width=0.5\textwidth]{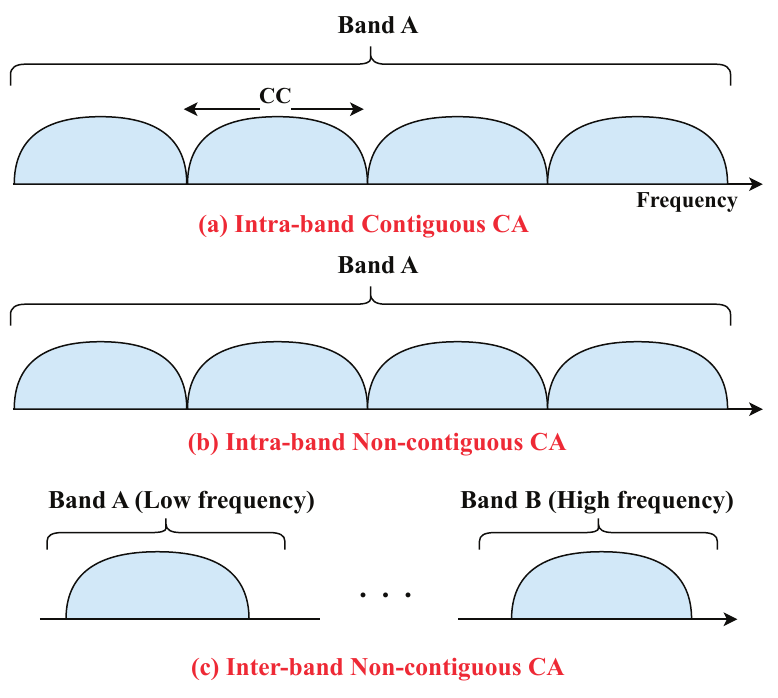}
	\caption{Three types of CA \cite{lee2013survey}.}
	\label{CA type}
\end{figure}

In general, there are three types of CA as shown in Fig. \ref{CA type},
namely intra-band contiguous CA, 
intra-band non-contiguous CA, 
and inter-band non-contiguous CA. 
The intra-band CA is equivalent to transmitting and receiving using larger points of Inverse Fast Fourier Transform (IFFT) and Fast Fourier Transform (FFT), 
so that the traditional Two-dimensional FFT (2D-FFT) algorithm is still applicable.
Towards 5G-A, the sub-6 GHz and millimeter wave (mmWave) spectrum bands are standardized.
Furthermore, the full-spectrum will be applied in 6G. 
Hence, the inter-band non-contiguous CA is widely applied since it realizes more
efficient utilization of fragmented spectrum resources via aggregating 
high and low-frequency bands compared with the other two types of CA. 
Motivated by this observation, 
we adopt inter-band non-contiguous CA to aggregate the 
fragmented spectrum resources in ISAC signal design and processing. 
In addition, different types of CA correspond to different architectures of transmitter and receiver. 
In terms of inter-band CA, 
according to 3GPP TR 36.912 standard \cite{3gpp2010nr}, 
a structure of multi-Radio Frequency (RF) chain is applied in this paper, 
as shown in Fig. \ref{system structure}.
However, the different subcarrier spacing in high and low-frequency bands 
makes it difficult to directly fuse the sensing information 
in high and low-frequency bands to improve the sensing performance.

The researches on CA techniques applied to ISAC-enabled mobile communication systems are divided into two categories, 
namely the CA-enabled communication and the CA-enabled radar sensing.  
In terms of CA-enabled communication, 
Kang \textit{et al.} \cite{kang2020dual} proposed a broadband high 
linearity drive amplifier that enables dual carrier aggregation and 
supports sub-GHz standards.
Ginzberg \textit{et al.} \cite{ginzberg2022carrier} proposed a digital linearisation scheme for a dual-channel discontinuous carrier aggregation transmitter which reduces the out-of-band distortion caused by CA.
In terms of CA-enabled radar sensing, the radar signal processing 
for multi-band Orthogonal Frequency Division Multiplexing (OFDM) radar is the most related area.
The stepped OFDM radar and frequency-agile stepped OFDM radar 
adopts the similar techniques in the intra-band non-contiguous CA.
Clemens \textit{et al.} \cite{pfeffer2015stepped} designed a stepped 
carrier OFDM signal, 
which combines 20 different frequency bands with 200 MHz bandwidth
to form a 4 GHz bandwidth, 
achieving high range resolution. 
However, this signal processing method is only suitable for the targets
with low mobility.
Benedikt \textit{et al.} \cite{schweizer2017stepped} designed an improved 
Discrete Fourier Transform (DFT) algorithm that obtains high range and velocity resolution for automotive applications by offsetting the phase errors from the range.
In \cite{lellouch2008ofdm}, 
the concept of frequency-agile stepping OFDM is introduced 
to obtain high range resolution with
limited bandwidth and gives a new Doppler processing.
Furthermore,
Knill \textit{et al.} \cite{knill2018high} presented a frequency-agile sparse OFDM radar signal processing 
method combined with Compressed Sensing (CS) technique to obtain the same sensing performance 
as equivalent OFDM signal on the continuous spectrum bands without increasing 
the sampling rate and bandwidth resources. 
However, the above studies do not involve 
the aggregation of high and low-frequency bands, 
where the differentiated subcarrier spacings bring 
a great challenge for the CA-enabled ISAC signal processing.

The other related research area of CA-based ISAC 
signal processing is multi-band radar,
which improves the sensing accuracy over fragmented frequency 
bands by reconstructing the data in blank frequency bands.
In 1974, Hogbom \textit{et al.} \cite{hogbom1974observations} proposed 
a deconvolution-based method to estimate blank band data and 
construct a complete two-dimensional (2D) radar image.
In 1992, Cuomo \cite{cuomo1992bandwidth} introduced the 
linear predictive Band Width Extrapolation (BWE) technique 
using an autoregressive time series model to extrapolate bandwidth 
at blanking positions to achieve high range resolution.
In 2004, Suwa \textit{et al.} \cite{suwa2004bandwidth} 
proposed a Polarisation Band Width Extrapolation (PBWE) technique based 
on polarised radar data that achieves higher resolution than the 
conventional BWE technique.
In 2009, Stoica \textit{et al.} \cite{stoica2009missing} applied 
the weighted least squares Iterative Adaptive Approach (IAA) in
radar sensing and proposed a missing data recovery approach 
applicable to arbitrary data missing patterns.
In 2015, Van Khanh \textit{et al.} \cite{VanKhanh2015} applied the 
BWE technique to the Linear Frequency Modulation (LFM) narrow band 
radar and explored the BWE technique based on Matched Filtering (MF), 
which results in improved range resolution.
In 2016, Zhang \textit{et al.} \cite{zhang2014coherent} presented a 
coherent signal processing and multi-band fusion algorithm 
based on a sparse signal model, which is adaptable to the case of 
large spectral spacing between multi-band signals.
However, the above studies are oriented for pulsed 
signal or LFM signal. 
The study on OFDM signal is very rare in the area of multi-band radar.

In this paper, we propose a CA-based staggered pilot structure for ISAC 
signal design. 
With CA techniques, 
the representative low and high-frequency bands, namely 5.9 GHz \cite{cheng2007mobile} and 24 GHz \cite{zhang200724ghz} 
frequency bands are selected as the fusion frequency bands. 
The parameters of OFDM signals are different in high and low-frequency bands, 
which brings challenge for the sensing information fusion in different frequency bands. 
Facing this challenge, 
the channel information matrix fusion method for CA-based staggered pilot ISAC signal 
is introduced. 
Then, a CS-based 2D-FFT algorithm is designed for ISAC signal processing.
Finally, the Cram\'{e}r-Rao Lower Bounds (CRLBs) 
for the proposed CA-based ISAC signal are derived. 
The detailed contributions are as follows.

\begin{itemize}
	\item[$\bullet$] \textbf{CA-based staggered pilot ISAC signal}: 
 	The staggered pilot structure is applied for CA-based ISAC signal design, 
 	where the block pilot is used in high-frequency band and the comb pilot is used 
 	in low-frequency band, which can comprehensively achieve better performance than other CA-based pilot signals.
\end{itemize}
\begin{itemize}
	\item[$\bullet$] \textbf{Channel information matrix fusion method}: 
	With the proposed CA-based staggered pilot ISAC signal, 
	the channel information matrix fusion method is proposed to 
	fuse the sensing information 
	in high and low-frequency bands with different parameters of OFDM signal 
	such as subcarrier spacing and length of OFDM symbols.
\end{itemize}
\begin{itemize}
	\item[$\bullet$] \textbf{CS-based 2D-FFT algorithm}: 
	The problem of 
	sidelobe elevation and sensing performance degradation caused by the 
	discontinuity of the modified channel information matrix is solved by 
	the proposed CS-based 2D-FFT algorithm, 
	which demonstrates strong anti-noise 
	capability in target sensing.
\end{itemize}
\begin{itemize}
	\item[$\bullet$] \textbf{CRLBs of CA-based ISAC Signal}: 
	The CRLBs for velocity 
	and range estimation under the proposed ISAC signal are derived. 
\end{itemize}

The rest of this paper is organized as follows. 
In Section \uppercase\expandafter{\romannumeral2}, 
the CA-based staggered pilot signal model is provided. 
In Section \uppercase\expandafter{\romannumeral3}, 
the ISAC signal processing methods are proposed. 
In Section \uppercase\expandafter{\romannumeral4}, 
the CRLBs for range and velocity estimation are revealed. 
In Section \uppercase\expandafter{\romannumeral5}, 
simulation results are presented. 
Finally, this paper is summarized in Section \uppercase\expandafter{\romannumeral6}. Table \ref{table_0} shows the abbreviations and notations used in this paper.

\section{Carrier Aggregation-based ISAC Signal}\label{s2}

\begin{figure}[!htbp]
	\centering
	\includegraphics[width=0.48\textwidth]{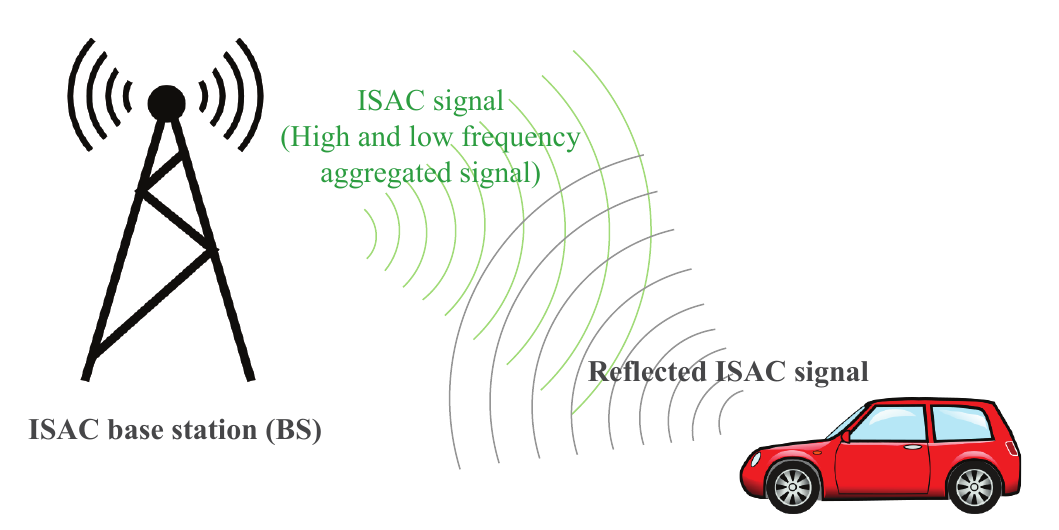}
	\caption{ISAC base station (BS) for vehicle communication and sensing.}
	\label{ISAC scenario}
\end{figure}

As shown in Fig. \ref{ISAC scenario}, 
the ISAC system for vehicle communication and sensing is considered. 
The OFDM signals on 5.9 GHz and 24 GHz frequency bands are representative frequency bands 
for vehicular communication \cite{1016299094.nh}. 
Hence, this paper adopts 5.9 GHz and 24 GHz frequency bands to 
design the CA-based ISAC signal. 
To make full use of spectrum resources and achieve highly accurate sensing performance, 
the CA-based staggered pilot ISAC signal is designed.

\subsection{The CA-based staggered pilot ISAC signal}

\begin{figure}[!h]
	\centering
	\includegraphics[width=0.5\textwidth]{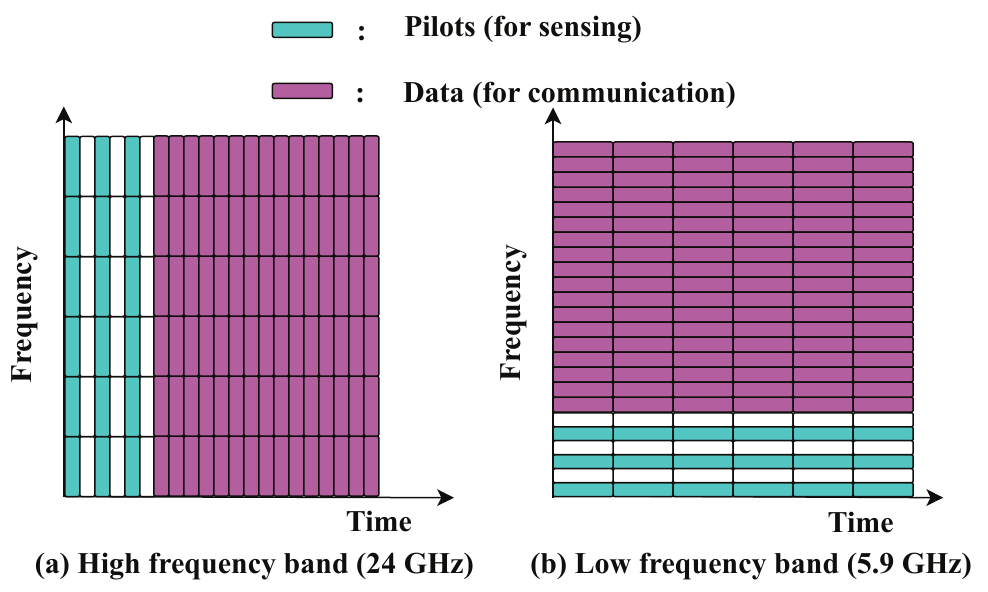}
	\caption{CA-based staggered pilot ISAC signal.}
	\label{ISAC signal}
\end{figure}

As shown in Fig. \ref{ISAC signal}, 
the structure of the CA-based staggered pilot ISAC signal is provided. 
Fig. \ref{ISAC signal}(a) represents the structure of block pilot in high-frequency band, 
which is discrete in time domain and continuous in frequency domain, 
so that the signal is insensitive to frequency-selective fading. 
Fig. \ref{ISAC signal}(b) shows the structure of comb pilot in low-frequency band, 
which is discrete in frequency domain and continuous in time domain. 
Therefore, the signal is insensitive to time-selective fading.

The comb pilot signal in low-frequency band at the transmitter (TX) 
is expressed as \cite{ozkaptan2018ofdm}
\begin{equation}\label{eq1}
	{x_1}\left( t \right) = \sum\limits_{m = 1}^{M} {\sum\limits_{n = 1}^{N'}} {{d_{\text{\tiny TX1}}}\left( {m,kn} \right)} \cdot \exp \left( {j2\pi {kn}\Delta ft} \right),
\end{equation}
where $N'= N/k$ is the number of subcarriers 
at the comb pilot location in each OFDM symbol,
$N$ is the total number of subcarriers in each OFDM symbol, 
and $ k $ is the interval of comb pilot. 
It is noted that $ k =\Delta f_2/\Delta f_1 $ is a pre-requisite for the range estimation, 
where $ \Delta f_1 $ and $ \Delta f_2 $ are the subcarrier spacings in low 
and high-frequency bands, respectively. 
Similarly, the time-domain expression of the block pilot signal 
in the high-frequency band is  \cite{1015730113.nh}
\begin{equation}\label{eq2}
	x_2\left( t \right) = \sum\limits_{m = 1}^{M'} {\sum\limits_{n = 1}^{N}} {{d_{\text{\tiny TX2}}}\left( {qm,n} \right)} \cdot \exp \left( {j2\pi {n}\Delta ft} \right),
\end{equation}
where $M' = M/q$ is the number of OFDM symbols at the block pilot location in a frame, 
$ M $ is the total number of OFDM symbols transmitted in one frame of data, 
and $ q $ is the interval of block pilot. 
According to \eqref{eq1} and \eqref{eq2}, 
the CA-based staggered pilot ISAC signal is expressed as
\begin{equation} \label{eq3}
	\begin{aligned}
		{x_{{\text{\tiny CA1}}}\left( t\right) } =& \sum\limits_{m = 1}^{M} {\sum\limits_{n = 1}^{N'}} {{d_{\text{\tiny TX1}}}\left( {m,kn} \right)} \cdot \exp \left( {j2\pi t \left( {kn\Delta {f_1}+{f_{c1}}}\right) } \right) \\
		&+ \sum\limits_{m = 1}^{M'} {\sum\limits_{n = 1}^{N}} {{d_{\text{\tiny TX2}}}\left( {qm,n} \right)} \cdot \exp \left(  {j2\pi t \left( {n\Delta {f_2}+{f_{c2}}}\right) } \right) ,	
	\end{aligned}
\end{equation}
where $d_{\text{\tiny TX1}} $ and  $d_{\text{\tiny TX2}} $ 
denote the modulation symbols in low and high-frequency bands, 
respectively. 
$ f_{c1} $ and  $ {f_{c2}} $ represent 
the carrier frequencies of the low and high-frequency bands, respectively.

The echo signal reflected from the target contains 
Doppler frequency shift $ {f_d} = 2v/\lambda = {2v{f_c}/{c_0}} $ and delay $ \tau = 2R/c_0 $, 
where $ R $ and $ v $ are the range and relative velocity of target, respectively, 
and $ \lambda $ is the wavelength of signal. 
Therefore, the echo signal at the receiver (RX) after down-conversion is expressed as
 \begin{equation} \label{eq4}
	\begin{aligned}	
		{y\left( t\right) }	=& \Bigl(\textbf{H}_1 \left( {m,kn}\right) \sum\limits_{m = 1}^{M}{\tiny {\sum\limits_{n = 1}^{N'}}} 
		{d_{\text{\tiny TX1}} \left( {m,kn} \right) } \cdot e^{j2 \pi nk \Delta f_1 \left( {t - \frac{2R}{c_0}} \right) }  \\
		&  \cdot { e^{ j2 \pi m T_1 \frac{2v f_{c1}}{c_0}}} 
		+ \textbf{H}_2 \left( {qm,n}\right) \sum\limits_{m = 1}^{M'}{\sum\limits_{n = 1}^{N}} 
		{d_{\text{\tiny TX2}} \left( {qm,n} \right) } \\ 
		&  \cdot e^{j2 \pi nk \Delta f_2 \left( {t - \frac{2R}{c_0}} \right) }  
		\cdot{ e^{ j2 \pi qm T_2 \frac{2v f_{c2}}{c_0}}} \Bigr) + w\left( t \right) ,
	\end{aligned}
\end{equation}
where $\textbf{H}_1\left( {m,n} \right)$ and $\textbf{H}_2\left( {m,n} \right)$ 
are the channel gain on the $n$-th subcarrier of the $m$-th OFDM symbol 
at high and low-frequency bands, 
respectively, which consist of channel fading and Radar Cross Section (RCS), 
$ w\left( t \right)  $ is the Additive White Gaussian Noise (AWGN). 
$ T_1 $ and $ {T_2} $ represent the total length of the OFDM symbols 
in low and high-frequency bands, respectively. 
The symbol length is $ T = 1/ \Delta f  + T_{CP} $, 
with $ T_{CP} $ representing the length of Cyclic Prefix (CP) 
and $ \Delta f $ representing the subcarrier spacing.

In addition to the above CA-based staggered pilot signal, 
there are other three aggregated structured signals, 
which are high-frequency comb and low-frequency block aggregated pilot signal in (\ref{eq5}), 
high and low-frequency full-block aggregated pilot signal in (\ref{eq6}), 
and high and low-frequency full-comb aggregated pilot signal in (\ref{eq7}), 
where the interval of comb pilot is $k$, and the interval of block pilot is $q$.
\begin{equation} \label{eq5}
	\begin{aligned}
		{x_{{\text{\tiny CA2}}}\left( t\right) } =& \sum\limits_{m = 1}^{M'} {\sum\limits_{n = 1}^{N}} {{d_{\text{\tiny TX1}}}\left( {qm,n} \right)} \cdot \exp \left( {j2\pi t \left( {n\Delta {f_1}+{f_{c1}}}\right) } \right) \\
		&+ \sum\limits_{m = 1}^{M} {\sum\limits_{n = 1}^{N'}} {{d_{\text{\tiny TX2}}}\left( {m,kn} \right)} \cdot \exp \left(  {j2\pi t \left( {kn\Delta {f_2}+{f_{c2}}}\right) } \right) ,		
	\end{aligned}
\end{equation}
\begin{equation} \label{eq6}
	\begin{aligned}
		{x_{{\text{\tiny CA3}}}\left( t\right) }= & \sum\limits_{m = 1}^{M'} {\sum\limits_{n = 1}^{N}} {{d_{\text{\tiny TX1}}}\left( {qm,n} \right)} \cdot \exp \left( {j2\pi t \left( {n\Delta {f_1}+{f_{c1}}}\right) } \right) \\
		&+ \sum\limits_{m = 1}^{M'} {\sum\limits_{n = 1}^{N}} {{d_{\text{\tiny TX2}}}\left( {qm,n} \right)} \cdot \exp \left(  {j2\pi t \left( {n\Delta {f_2}+{f_{c2}}}\right) } \right) ,	
	\end{aligned}
\end{equation}
\begin{equation} \label{eq7}
	\begin{aligned}
		{x_{{\text{\tiny CA4}}}\left( t\right) } =& \sum\limits_{m = 1}^{M} {\sum\limits_{n = 1}^{N'}} {{d_{\text{\tiny TX1}}}\left( {m,kn} \right)} \cdot \exp \left( {j2\pi t \left( {kn\Delta {f_1}+{f_{c1}}}\right) } \right) \\
		&+ \sum\limits_{m = 1}^{M} {\sum\limits_{n = 1}^{N'}} {{d_{\text{\tiny TX2}}}\left( {m,kn} \right)} \cdot \exp \left(  {j2\pi t \left( {kn\Delta {f_2}+{f_{c2}}}\right) } \right) .		
	\end{aligned}
\end{equation}

\subsection{Framework of ISAC signal processing}

\begin{figure*}[!htbp]
	\centering
	\includegraphics[width=0.95\textwidth]{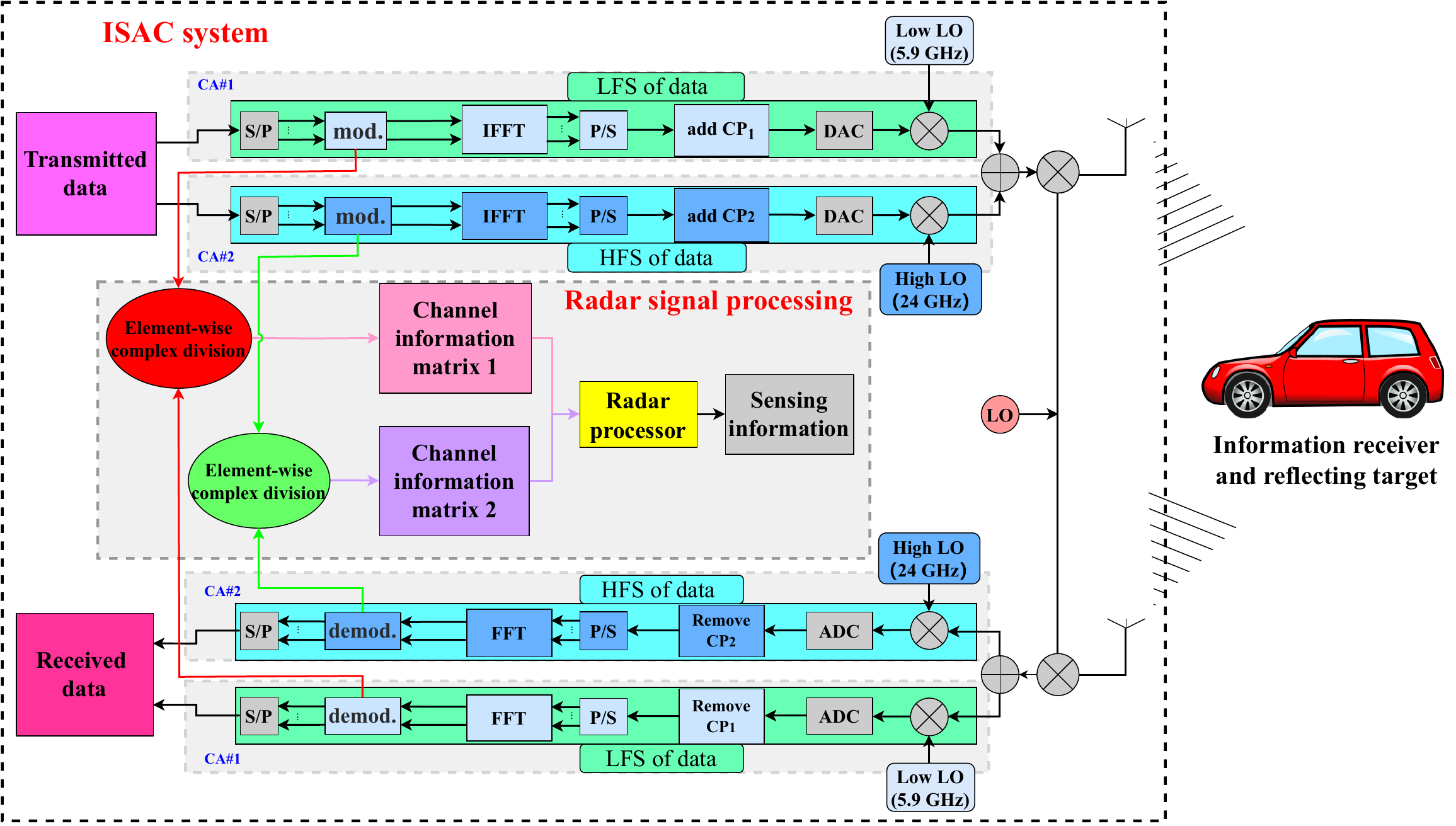}
	\caption{CA-based ISAC signal processing, with the abbreviations HFS: high-frequency stream, LFS: low-frequency stream, LO: local oscillator, CP: cyclic prefix, FFT: fast Fourier transform, IFFT: inverse fast Fourier transform, ADC: analog-to-digital converter, DAC: digital-to-analog converter, S/P: serial-to-parallel conversion, P/S: parallel-to-serial conversion. }
	\label{system structure}
\end{figure*}

\subsubsection{Signal processing at TX}
As shown in Fig. \ref{system structure}, 
the transmitted data is divided into two data streams: 
High-frequency Stream (HFS) and Low-frequency Stream (LFS), which are also denoted by high-frequency CC and low-frequency CC, respectively.
In the HFS, 
the transmitted data is modulated to form a modulated symbol. 
Undergoing block-wise inverse fast Fourier transform, 
subsequent parallel-to-serial conversion and the addition of CP, 
the signal is converted to an analog signal and mixed with the high-frequency LO 
signal to provide a high-frequency analog signal \cite{sturm2011waveform}.
Compared with the HFS, 
the LFS is mixed with the low-frequency LO to obtain low-frequency analog signal. 
Then, the low-frequency analog signal and high-frequency analog signal are 
combined and fed into a single antenna \cite{3gpp2010nr}. 
When the multiple antennas are applied, 
we need to consider precoding schemes to improve the 
performance of communication and sensing. 
Since a single antenna is applied in this paper, 
the problems associated with multiple antennas are not considered.

\subsubsection{Signal processing at RX}
For communication, the demodulation process is similar to traditional OFDM demodulation \cite{thompson2008constant} since the data stream has been separated into HFS and LFS at RX.
At the RX, 
the same steps are performed in reverse order. 
The received modulated symbol is recovered by FFT. 
Then, we can also use the traditional maximum likelihood detection 
to obtain the communication data.
For sensing,
the modulated symbols at RX are divided by the modulated symbols at TX to 
obtain the channel information matrix, 
which is used for radar sensing. 
A detailed description of the ISAC signal processing is provided in Section \uppercase\expandafter{\romannumeral3}.

The differences in the subcarrier spacings and 
the length of OFDM symbols in high and low-frequency bands 
bring challenges for the processing of the CA-based ISAC signal processing, 
which will be solved in Section \uppercase\expandafter{\romannumeral3}.

\section{ISAC Signal Processing}\label{s3}

In terms of radar signal processing for OFDM-based ISAC signal, 
Sturm \textit{et al.} \cite{sturm2011waveform} proposed 2D-FFT algorithm,
which takes full advantage of the 2D structure of OFDM signal in 
efficient radar sensing \cite{wei2023integrated}. 
Zuo \textit{et al.} 
\cite{DZYX202012019} proposed a CS-based radar signal processing for 
index-modulated OFDM (OFDM-IM) signals. 
Based on these previous approaches, 
we propose a CA-based ISAC signal processing algorithm in this section.

\begin{figure}[!htbp]
	\centering
	\includegraphics[width=0.46\textwidth]{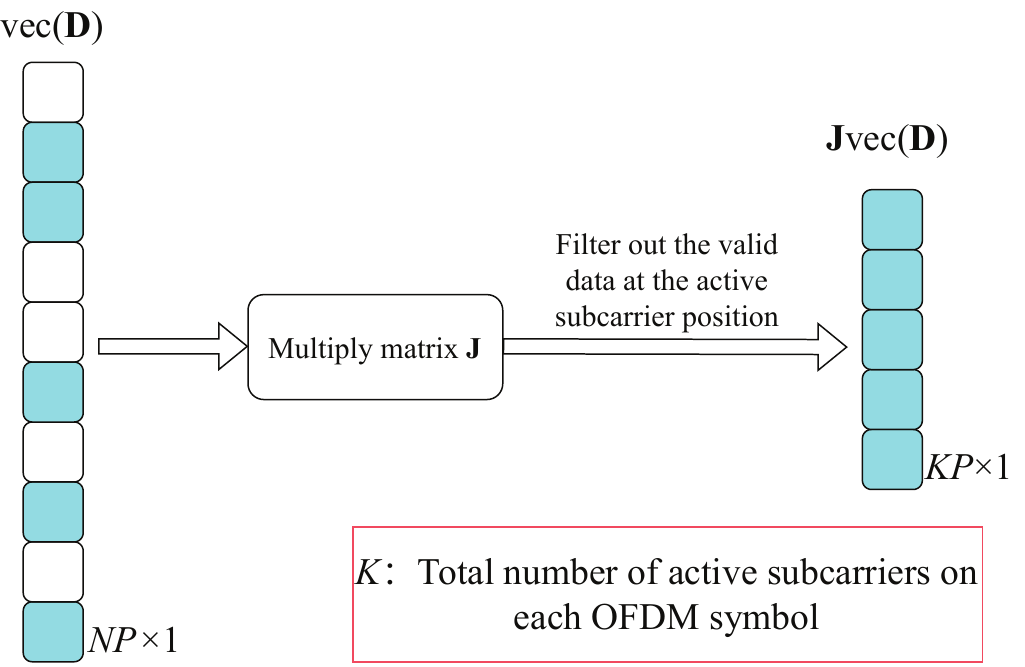}
	\caption{filtering process using selection matrix \textbf{J}.}
	\label{filtering process}
\end{figure}

The CS methods apply the sparsity of echo signal 
to reconstruct the incompletely sampled 
channel information matrix in 2D-FFT method, 
thus reducing the Fourier sidelobe deterioration. 
The channel information matrix is expressed as
\begin{equation}\label{eq8}
	\textbf{D} =\textbf{F}_N \textbf{X} \textbf{F}_P^\mathrm{H} + \bar{\textbf{W}},
\end{equation}
where $ \textbf{X} $ represents range-velocity profile of target, 
$ \textbf{F}_N \in \mathbb{C}^{N \times N} $ and $ \textbf{F}_P \in \mathbb{C}^{P \times P} $
are Fourier transform matrices, respectively. 
$ N $  is the total number of subcarriers, 
$ P $ is the number of OFDM symbols, 
$ \left( \cdot\right) ^{\mathrm{H}} $ is the conjugate transpose of the matrix, 
$  \bar{\textbf{W}}  $ is AWGN matrix. 
The selection matrix
$ \textbf{J} \in \mathbb{C}^{KP \times NP}  $  is defined to 
filter out valid data based on subcarrier activation. 
The filtering process is shown in Fig. \ref{filtering process}.
Multiplying both sides of \eqref{eq8} by the matrix $ \textbf{J} $, we have
\begin{equation}\label{eq9}
	\textbf{d} =\textbf{A} \textbf{x} + \textbf{w},
\end{equation}
where
\begin{equation}\label{eq10}
	\begin{cases}
		{ \textbf{d} = \textbf{J} \mathrm{{vec}} \left( \textbf{D} \right) } \\
		{ \textbf{A} =\textbf{J} \left( \textbf{F}_P^\mathrm{H} \otimes \textbf{F}_N \right)} \\
		{\textbf{x} = \mathrm{{vec}} \left( \textbf{X} \right)  } \\
		{ \textbf{w} =\textbf{J} \mathrm{{vec}} \left(  \bar{\textbf{W}} \right)  }
	\end{cases}
\end{equation}
In (\ref{eq10}), 
$\mathrm{vec} \left( \cdot \right)$ represents stacking matrix in columns, 
$ \otimes $ refers to Kronecker product, 
$\textbf{A}$ is the sensing matrix in CS. 
When $ \textbf{x} $ is a sparse vector and  $ \textbf{A} $ satisfies the 
Restricted Isometry Property (RIP) condition \cite{davenport2012introduction}, 
the sparse signals can be reconstructed. 
Since the solution of \eqref{eq9} is NP-hard, 
\eqref{eq9} is transformed into \eqref{eq11}.
\begin{equation} \label{eq11}
	\hat{\textbf{x}} =\arg\mathop{\min}_{\textbf{x}} \frac{1}{2} \lVert { \textbf{d} - \textbf{A} \textbf{x} } \rVert_2^2 + \lambda \lVert \textbf{x} \rVert_1 .
\end{equation}

Problem \eqref{eq11} is a least square optimization problem 
that can be solved by the Orthogonal Matching Pursuit (OMP) algorithm. OMP is an iterative greedy algorithm that is utilized to recover high-dimensional sparse signals from a limited number of noisy linear measurements. During each iteration, OMP selects the column that exhibits the highest correlation with the current residual and incorporates it into the set of chosen columns. 
Detailed descriptions of OMP can be found in \cite{cai2011orthogonal}. 
By solving \eqref{eq11}, 
the range-velocity profile is obtained for the complete OFDM signal, 
avoiding the effect of Fourier sidelobe deterioration.

In this section, 
the CA-based ISAC signal processing algorithm is proposed, 
which takes full advantage of the two channel information matrices in high and 
low-frequency bands, resulting in better sensing performance.

The CA-based staggered pilots signal adopts comb pilot in low-frequency band 
with interval of $ K = \Delta f_2/\Delta f_1 $ 
and block pilot in high-frequency band with interval of $ Q $. 
According to Fig. \ref{system structure}, 
when the low-frequency and high-frequency data streams pass through the DAC, 
they are mixed with high and low-frequency LOs  
to obtain high and low-frequency analog signals. 
At RX, the mixed high and low-frequency analog signals are separated by matched filtering. 
Using element-wise complex division, 
the channel information matrix $ \textbf{D}_{cf1} \in \mathbb{C}^{N \times M} $ in low-frequency band and the channel information matrix $ \textbf{D}_{cf2} \in \mathbb{C}^{N \times M} $ in high-frequency band are obtained at the RX.
\begin{equation}\label{eq12}
	\begin{aligned}
		&\textbf{D}_{cf1} = \textbf{H}_1 \left( {m,n} \right) \circ \left( {k}_{r1} \otimes {k}_{d1}  \right), \\
		&\textbf{D}_{cf2} = \textbf{H}_2 \left( {m,n} \right) \circ \left( {k}_{r2} \otimes {k}_{d2}  \right),
	\end{aligned}
\end{equation}
where $\circ$ denotes Hadamard product.
Combining the structures of high-frequency block pilot and low-frequency comb pilot, we have
\begin{equation}\label{eq13}
	\begin{aligned}
		&k_{d1}\left( m \right) = \exp\left( j2\pi m T_1 \frac{2v_0 f_{c1}}{c_0} \right),  m= 0,1,2,...,M-1, \\
		&k_{r1}\left( n \right) = \exp\left( -j2\pi n \Delta f_1 \frac{2R}{c_0} \right), n= 0,K,2K,...,aK, \\
	\end{aligned}
\end{equation}
where $aK \le N-1$. And we have
\begin{equation}\label{eq14}
	\begin{aligned}
		&k_{d2}\left( m \right) = \exp\left( j2\pi m T_2 \frac{2v_0 f_{c2}}{c_0} \right),  m= 0,Q,2Q,...,bQ, \\
		&k_{r2}\left( n \right) = \exp\left( -j2\pi n \Delta f_2 \frac{2R}{c_0} \right), n= 0,1,2,...,N-1, 
	\end{aligned}
\end{equation}
where $bQ \le M-1$. 
In order to represent the structural form of the two channel information matrices clearly, 
the matrices are rewritten as (\ref{eq15}) and (\ref{eq16}), 
where the elements in $\textbf{D}_{cf1}$
and $\textbf{D}_{cf2}$ denote the elements after the division of the received modulation symbols by the transmitted modulation symbols.
\begin{figure*}
\centering
\begin{minipage}{.48\linewidth}
\begin{equation}\label{eq15}
	\textbf{D}_{cf1} = \left[
	\begin{array}{ccccc}
		z_{0,0} & z_{0,1}  & z_{0,2}  & \dots  & z_{0,M-1}  \\
		0& 0 & 0 & 0 & 0 \\
		\vdots&\vdots  &\vdots  &\ddots  &\vdots  \\
		0& 0 & 0 & \dots &0  \\
		z_{K,0} & z_{K,1}  &z_{K,2}  &\dots  &z_{K,M-1}  \\
		0& 0 & 0 & \dots & 0 \\
		\vdots&\vdots  &\vdots  &\ddots  &\vdots  \\
		0& 0 & 0 & \dots &0  \\
		z_{2K,0}&z_{2K,1}  &z_{2K,2}  &\dots  &z_{2K,M-1}  \\
		0& 0 &0  & \dots & 0 \\
		\vdots&\vdots  &\vdots  &\ddots  &\vdots  \\
		0& 0 & 0 & \dots & 0 \\
		\vdots&\vdots  &\vdots  &\ddots  &\vdots  \\
		z_{aK,0}&z_{aK,1}  & z_{aK,2} & \dots & z_{aK,M-1} \\
		0&  0& 0 & \dots & 0 \\
		\vdots& \vdots &  \vdots& \ddots &\vdots 
	\end{array}
	\right]  ,
\end{equation}
\end{minipage}
\hfill
\begin{minipage}{.48\linewidth}
\begin{equation}\label{eq16}
	\textbf{D}_{cf2} = \left[
	\begin{array}{ccccc}
		s_{0,0}&s_{1,0}  &s_{2,0}  &\dots  &s_{N-1,0}  \\
		0&  0&  0&  \dots&0  \\
		\vdots& \vdots & \vdots & \ddots &\vdots  \\
		0&  0&  0& \dots & 0 \\
		s_{0,Q}& s_{1,Q}&s_{2,Q}  & \dots  &s_{N-1,Q}  \\
		0&  0&  0&  \dots&0  \\
		\vdots& \vdots & \vdots & \ddots &\vdots  \\
		0&  0&  0& \dots & 0 \\
		s_{0,2Q}&s_{1,2Q}  &s_{2,2Q}  &\dots  & s_{N-1,2Q} \\
		0&  0&  0&  \dots&0  \\
		\vdots& \vdots & \vdots & \ddots &\vdots  \\
		0&  0&  0& \dots & 0 \\
		\vdots& \vdots & \vdots & \ddots &\vdots  \\
		s_{0,bQ}&s_{1,bQ}  &s_{2,bQ}  &\dots  & s_{N-1,bQ} \\
		0&  0&  0&  \dots&0  \\
		\vdots& \vdots & \vdots & \ddots &\vdots 
	\end{array}
	\right]^\mathrm{T}.
\end{equation}
\end{minipage}
	{\noindent} \rule[-10pt]{18cm}{0.05em}
\end{figure*}

\subsection{Range Estimation Method}

The different subcarrier spacings in high and low-frequency bands bring
challenges for the range estimation using CA-based ISAC signal. 
According to 2D-FFT algorithm, 
the column vector of channel information matrix is applied 
in calculating the range of target \cite{sturm2011waveform}.

\subsubsection{Fusion of channel information matrices}

Comparing \eqref{eq13} and \eqref{eq14}, 
it is discovered that the different parameters between $k_{r1}\left( n \right)$ and
$ k_{r2}\left( n \right) $ are $ n $ and $ \Delta f $. 
As mentioned in Section II, 
$ \Delta f_2 = K\Delta f_1 $. 
Replacing the variable $ n $ in $ k_{r1}\left( n \right) $ with $ Kn' $, 
the transformation of $ k_{r1}\left( n \right) $ is
\begin{equation}\label{eq17}
	k_{r1}\left( n'\right) = \exp\left( {-j2 \pi n' \left( K\Delta f_1 \right) \frac{2R}{c_0} } \right), \; n'=0,1,2,...,a,
\end{equation}
where $ K $  is the interval of comb pilot. 
Comparing $k_{r1}\left( n'\right)$ in (\ref{eq17}) 
and $k_{r2}\left( n \right)$ in (\ref{eq14}), 
$k_{r1}\left( n'\right)$ is a subsequence of the preceding $ a+1 $  elements of $ k_{r2}\left( n \right)$.

\subsubsection{CS-based range estimation algorithm}
In order to estimate the range of target, 
the modulation symbols in high and low-frequency bands at the RX are recovered. 
Then, the channel information matrix $ \textbf{D}_{cf2} $ in high-frequency band
and the channel information matrix $ \textbf{D}_{cf1} $ in low-frequency band
are obtained. 
The elements of $ \textbf{D}_{cf1} $ in 
the positions of pilot's subcarriers are rearranged, 
and the channel information matrix  $ \textbf{D}^{'}_{cf1} $ is obtained.
Finally, in order to overcome the impact of non-continuous subcarriers or symbols 
on the performance of sensing, 
the CS-based target estimation method 
in \cite{DZYX202012019} is applied, 
which has been introduced at the beginning of Section \ref{s3}. 
The sensing matrix in \cite{DZYX202012019} needs to be adjusted for non-continuous situations. 
The process of adjusting the sensing matrices used in this paper 
is described below.
Beck \textit{et al.} proposed the fast iterative shrinkage-thresholding algorithm (FISTA) \cite{beck2009fast}, 
which applies the gradient descent to solve optimization problems.
FISTA achieves fast convergence by generalizing the iterative idea of Nesterov’s method \cite{nesterov1983method}. 
Thus, FISTA is used instead of the OMP reconstruction algorithm to 
obtain fast convergence speed.
The above algorithm is referred to as CS-IDFT. 
When processing the channel information matrix $\textbf{D}^{'}_{cf1}$, 
a column in the channel information matrix is selected to obtain the sensing matrix 
based on the location of the valid data in a column. 
Define a selection matrix $ \textbf{Q} \in \mathbb{C}^{N \times N} $ as follows, 
where the elements in the first $a+1$ rows are $ 1 $ and the rest elements are $ 0 $.
\begin{equation}\label{eq18}
	\textbf{Q} = \left[
	\begin{array}{ccccc}
		1_{0,0}	& 1_{0,1}  & 1_{0,2} & \dots&1_{0,N-1} \\
		\vdots& \vdots &\vdots  &\ddots& \vdots  \\
		1_{a,0}	&1_{a,1} & 1_{a,2} & \dots&1_{a,N-1} \\
		0&  0& 0 &\dots&0  \\
		\vdots& \vdots &\vdots  &\ddots&\vdots \\
		0 & 0 & 0 &\dots&0 \\
	\end{array}
	\right].
\end{equation}

\begin{figure}[!h]
	\centering
	\includegraphics[width=0.45\textwidth]{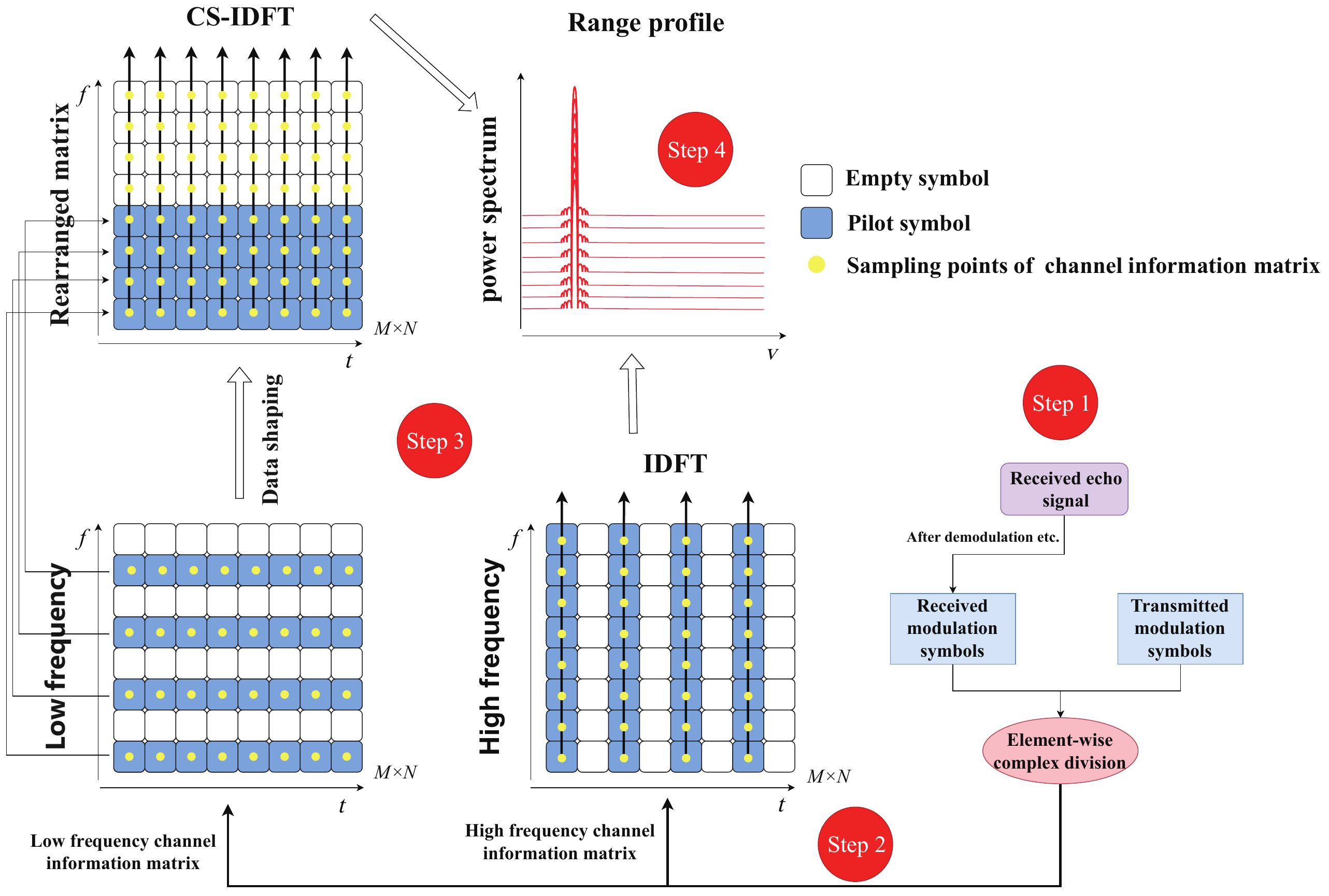}
	\caption{Signal processing for range estimation.}
	\label{range method}
\end{figure}

The sensing matrix in CS is represented as the product of the sparse basis and the measurement matrix. 
The channel information matrix can be transformed to the range-Doppler domain 
under the discrete Fourier transform basis, 
while the radar signals are generally sparse in the range-velocity domain \cite{knill2018high}. 
Therefore, the sensing matrix in this paper takes the form of Fourier transform basis. 
$ \psi \in \mathbb{C}^{N \times N} $  
is the discrete inverse Fourier matrix.
Hence, the modified sensing matrix is
$ \textbf{Q} \circ \psi^{-1} $.
For the remaining columns, 
the sensing matrix is obtained as (\ref{eq18}). 
Using the sensing matrix and the data in each column of $\textbf{D}^{'}_{cf1}$, 
we can obtain power spectrum of range by solving (\ref{eq11}). 
The range estimation algorithm is shown in \hyperref[tab1]{Algorithm 1},
which is explained intuitively in Fig. \ref{range method}. 
The channel information matrix in low-frequency band is 
firstly processed to obtain the rearranged matrix $\textbf{D}^{'}_{cf1}$. 
Then, the power spectrum of range is obtained using the CS-IDFT algorithm for 
$\textbf{D}^{'}_{cf1}$.
The power spectrum of range is obtained using the IDFT of the 
channel information matrix in high-frequency band. 
The power spectra of range in high and low-frequency bands are superimposed. 
Finally, the index value $ ind_n $ is obtained by searching the peak of 
the superimposed power spectrum
and the estimated range of target is
\begin{equation}\label{eq19}
	R = ( \lfloor ind_n \rfloor c_0 )/ (2\Delta f_2 N ) .
\end{equation}

\begin{table}[!ht]
\centering
\label{tab1}
\resizebox{0.48\textwidth}{!}{
\setlength{\arrayrulewidth}{0.9pt}
\begin{tabular}{rllll}
\hline
\multicolumn{5}{l}{\textbf{Algorithm 1:} Range Estimation Algorithm}  \\ \hline
\multirow{-8}{*}{\textbf{Input:} }        & \multicolumn{4}{l}{\begin{tabular}[c]{@{}l@{}}Channel information matrix $\textbf{D}_{cf1} $ in low-frequency\\ band; \\ Channel information matrix $ \textbf{D}_{cf2}$ in high-frequency\\ band; \\ Sensing matrix $ \textbf{Q} \circ \psi^{-1} $;\\ The intervals of comb and block pilots: $ K $ and $ Q $;\\ The number of OFDM symbols $ M $;\\ The number of subcarriers $ N $;\end{tabular}} \\
1:                      & \multicolumn{4}{l}{$\textbf{For}$ the $j$ row vector of $\textbf{D}_{cf1}$ in steps $K$ $\textbf{do}$} \\
2:                      & \multicolumn{4}{l}{$\hspace{0.5em}$ Initialize a zero matrix $ \textbf{D}^{'}_{cf1} \in \mathbb{C}^{N \times M}$ and $i=1$;}                                         \\
3:                      & \multicolumn{4}{l}{$\hspace{0.5em}$ Deposit the $j$ row of $\textbf{D}_{cf1}$ into $i$-th row of $\textbf{D}^{'}_{cf1}$;}                           \\
4:                      & \multicolumn{4}{l}{$\hspace{0.5em}$ $i=i+1$;}              \\
5:                      & \multicolumn{4}{l}{$\textbf{End For}$}               \\
6:                      & \multicolumn{4}{l}{Initialize a power spectrum of range $R_{r}$ and $\varpi=0$;}         \\
7:                      & \multicolumn{4}{l}{$\textbf{For}$ the $j$ column vector of $\textbf{D}_{cf2}$ in steps $Q$ $\textbf{do}$}           \\
\multirow{-2}{*}{8:}                     & \multicolumn{4}{l}{\begin{tabular}[c]{@{}l@{}}$\hspace{0.5em}$ The result of the IFFT operation on the $j$-th column  \\ $\hspace{0.5em}$ of $\textbf{D}_{cf2}$ is assigned to $\varpi$;\end{tabular}} \\ 
9:                      & \multicolumn{4}{l}{$\hspace{0.5em}$ $R_{r}=R_{r}+\|\varpi\|$;}   \\
10:                     & \multicolumn{4}{l}{$\textbf{End For}$}    \\
11:                     & \multicolumn{4}{l}{$\textbf{For}$ each column vector of $\textbf{D}^{'}_{cf1}$ $\textbf{do}$} \\
\multirow{-2}{*}{12:}                    & \multicolumn{4}{l}{\begin{tabular}[c]{@{}l@{}}$\hspace{0.5em}$ Input $ \textbf{Q} \circ \psi^{-1} $ and the column vector of $\textbf{D}^{'}_{cf1}$ \\ $\hspace{0.5em}$ to FISTA \cite{beck2009fast};\end{tabular}} \\
13:                     & \multicolumn{4}{l}{$\hspace{0.5em}$ The result of output of the FISTA is assigned to $\varpi$;} \\
14:                     & \multicolumn{4}{l}{$\hspace{0.5em}$ $R_{r}=R_{r}+\|\varpi\|$;}   \\
15:                     & \multicolumn{4}{l}{$\textbf{End For}$}               \\
16:                     & \multicolumn{4}{l}{Normalize $R_{r}$ and perform a peak search on it;}   \\
\multirow{-2}{*}{17:}                    & \multicolumn{4}{l}{\begin{tabular}[c]{@{}l@{}}Substitute the index value corresponding to \\ the searched peak into (\ref{eq19});\end{tabular}} \\
\textbf{Output:}        & \multicolumn{4}{l}{The estimation of the range $\hat{R}$}         \\ 
\hline
\end{tabular}}
\end{table}

\subsection{Velocity Estimation Method}

In the estimation of target's velocity, 
the difference in carrier frequencies in high and low-frequency bands leads to different resolutions of 
the velocity estimation, 
which brings challenges for the fusion of sensing information 
in high and low-frequency bands.

\subsubsection{Fusion of channel information matrices}

The relative velocity of target introduces a linear phase shift along the time axis, 
so that only the row vectors of channel information matrix need to be considered.

By observing $ k_{d1}\left( m \right)  $ in \eqref{eq13} and $ k_{d2}\left( m \right) $ in  \eqref{eq14}, 
it is discovered that the different parameters are $ m $, $ T $ and $ f_c $. 
As revealed in Fig. \ref{system structure}, 
the data streams of HFS and LFS use different modules of adding CP, 
which can dynamically adjust CP length. 
On the premise of meeting the maximum ranging, $ T_1 f_{c1} = T_2 f_{c2} $ can be satisfied by adding CP with different lengths. 
According to 2D-FFT method \cite{sturm2011waveform}, $ ind_m = \lfloor {(2v_0 f_c TM)/c_0 } \rfloor $ is obtained. 
Therefore, when the FFT is performed on the row vectors of high and low-frequency channel information matrices, 
the peak index values will be the same with $ T_1 f_{c1} = T_2 f_{c2} $. 
Therefore, the velocity of target can be estimated by the accumulation of high and low-frequency channel information matrices.

\subsubsection{CS-based velocity estimation algorithm}

\begin{figure}[!h]
	\centering
	\includegraphics[width=0.46\textwidth]{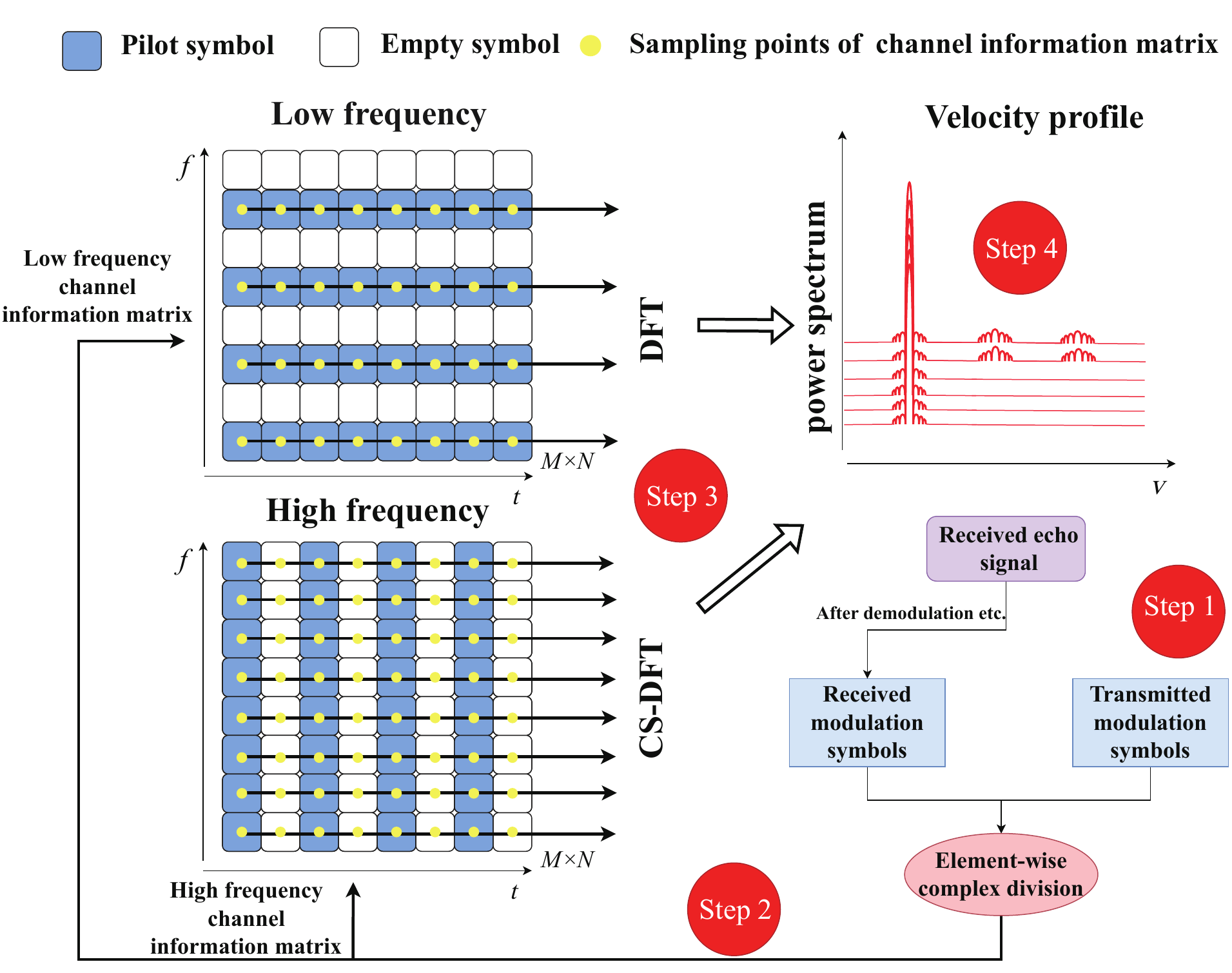}
	\caption{Signal processing for velocity estimation.}
	\label{velocity method}
\end{figure}

By observing $ k_{d2}\left( m \right)  $ in (\ref{eq14}), 
it is revealed that the row vector contains the null modulation symbol, 
and Fourier transform will cause the increase of sidelobe. 
According to \cite{1021843429.nh}, sidelobe and noise can be reduced by the estimation method based on CS. The FISTA reconstruction algorithm is used instead of the OMP algorithm. 
The algorithm here is refered to CS-DFT. 
When processing the channel information matrix $ \textbf{D}_{cf2} $, 
the sensing matrix for the first row vector in $ \textbf{D}_{cf2} $ is obtained based on the location of the valid data in the first row of $ \textbf{D}_{cf2} $. 
A selection matrix $ \textbf{P} \in \mathbb{C}^{M \times M} $ is defined based on the interval of block pilot $ Q $  and the number of OFDM symbols $ M $ 
\begin{equation}\label{eq20}
	\textbf{P} = \left[
	\begin{array}{cccc}
		1	& 1 & \dots  & 1 \\
		\vdots	& \vdots & \dots & \vdots \\
		1	& 1 & \dots  & 1 \\
		\vdots	& \vdots & \dots & \vdots 
	\end{array}
	\right],
\end{equation}
where each column of $\textbf{P}$ is 
\begin{equation}
	{\large p}_{ 1 \times M} =\left[1, 0_{1 \times (Q-1)},1, 0_{1 \times (Q-1)},...\right]^{\mathrm{T}} .
\end{equation}
Then the sensing matrix  $ \textbf{P} \circ \psi $ is obtained. 
For the remaining rows, the sensing matrix is obtained as (\ref{eq20}). 
Using the sensing matrix and the data in each row of $ \textbf{D}_{cf2} $, 
the power spectrum of velocity is obtained by solving \eqref{eq11}. 
The velocity estimation algorithm is shown in \hyperref[tab2]{Algorithm 2},
which is explained intuitively in Fig. \ref{velocity method}. 
The power spectrum of velocity is obtained by using the CS-DFT 
algorithm for the high-frequency channel information matrix. 
Then, the DFT is performed on the low-frequency channel information matrix 
according to the pilot position to obtain the power spectrum of velocity. 
The power spectra of velocity in high and low-frequency bands are superimposed. 
Finally, the index value $ ind_m $ is obtained by searching the peak of 
the superimposed power spectrum and the estimated velocity of target is 
\begin{equation}\label{eq21}
	v_0 = ( \lfloor ind_m \rfloor c_0 )/ (2 f_{c2}T_2 M ).
\end{equation}

\begin{table}[!ht]
\centering
\label{tab2}
\resizebox{0.48\textwidth}{!}{
\setlength{\arrayrulewidth}{1pt}
\begin{tabular}{rllll}
\hline
\multicolumn{5}{l}{\textbf{Algorithm 2:} Velocity Estimation Algorithm} \\ \hline
\multirow{-8}{*}{\textbf{Input:} }          & \multicolumn{4}{l}{\begin{tabular}[c]{@{}l@{}}Channel information matrix $\textbf{D}_{cf1} $ in low-frequency \\ band; \\ Channel information matrix $ \textbf{D}_{cf2}$ in high-frequency \\ band; \\ Sensing matrix $ \textbf{P} \circ \psi $;\\ The intervals of comb and block pilots: $ K $ and $ Q $;\\ The number of OFDM symbols $ M $;\\ The number of subcarriers $ N $;\end{tabular}} \\
1:                      & \multicolumn{4}{l}{Initialize a power spectrum of velocity $R_{v}$ and $\varpi=0$;} \\
2:                      & \multicolumn{4}{l}{$\textbf{For}$ the $j$-th row vector of $\textbf{D}_{cf1}$ in steps $K$ $\textbf{do}$}  \\
\multirow{-2}{*}{3:}                     & \multicolumn{4}{l}{\begin{tabular}[c]{@{}l@{}}$\hspace{0.5em}$ The result of the FFT operation for the $j$-th row \\ $\hspace{0.5em}$ of $\textbf{D}_{cf1}$ is assigned to $\varpi$;\end{tabular}}  \\
4:                      & \multicolumn{4}{l}{$\hspace{0.5em}$ $R_{v}=R_{v}+\|\varpi\|$;}  \\
5:                     & \multicolumn{4}{l}{$\textbf{End For}$}  \\
6:                     & \multicolumn{4}{l}{$\textbf{For}$ each row vector of $\textbf{D}_{cf2}$ $\textbf{do}$}  \\
\multirow{-2}{*}{7:}                     & \multicolumn{4}{l}{\begin{tabular}[c]{@{}l@{}}$\hspace{0.5em}$ Input $ \textbf{P} \circ \psi $ and the row vector of $\textbf{D}_{cf2}$ \\ $\hspace{0.5em}$ to FISTA \cite{beck2009fast};\end{tabular}}  \\
8:                     & \multicolumn{4}{l}{$\hspace{0.5em}$ The result of output of the FISTA is assigned to $\varpi$;} \\
9:                     & \multicolumn{4}{l}{$\hspace{0.5em}$ $R_{v}=R_{v}+\|\varpi\|$;}   \\
10:                    & \multicolumn{4}{l}{$\textbf{End For}$}               \\
11:                     & \multicolumn{4}{l}{Normalize $R_{v}$ and perform a peak search on it;}  \\
\multirow{-2}{*}{12:}                     & \multicolumn{4}{l}{\begin{tabular}[c]{@{}l@{}}Substitute the index value corresponding to \\ the searched peak into (\ref{eq21});\end{tabular}}  \\
\textbf{Output:}        & \multicolumn{4}{l}{The estimation of the range $\hat{v_0}$} \\ 
\hline
\end{tabular}}
\end{table}

\section{Performance of CA-based ISAC Signal}

In this section, the sensing performance of CA-based ISAC signals is analyzed using 
Root Mean Square Error (RMSE) and CRLB. 
Firstly, the ISAC signal processing algorithms under different pilot structures are provided.
Then, the CRLBs of ISAC signals under different pilot structures are derived.

\subsection{Other Types of Pilot Structures}

The above CA-enabled ISAC signal design adopts the pilot structure with 
block pilot in high-frequency band and comb pilot in low-frequency band. 
However, there are three other pilot structures, 
which are high-frequency comb and low-frequency block aggregated pilot, 
high and low-frequency full-block aggregated pilot, and high and low-frequency full-comb aggregated pilot.
The signal processing procedures under 
the other three pilot structures are as follows. 

\subsubsection{High-frequency comb and low-frequency block aggregrated pilot} 
The steps of signal processing are as follows.
\begin{itemize}
    \item \textbf{Step 1:} The low-frequency channel information 
    matrix $\mathbf{D}_{cf3}$ and 
    the high-frequency channel information matrix $\mathbf{D}_{cf4}$ are obtained.
    \item \textbf{Step 2:} We perform IDFT on the column vectors corresponding to the block pilot positions of $\mathbf{D}_{cf3}$, sum the results, search the power spectrum for the peak index value, and combine it with (\ref{eq19}) to get the range estimate $r_1$. Likewise, we perform CS-DFT on every row vector of $\mathbf{D}_{cf3}$, sum the results, combine the peak index value with (\ref{eq21}) to get the velocity estimation $v_1$.
    \item \textbf{Step 3:} We perform CS-IDFT on every column vector of $\mathbf{D}_{cf4}$, sum the results, search the power spectrum for the peak index value, and combine it with (\ref{eq19}) to get the range estimation $r_2$. Likewise, we perform DFT on the row vectors corresponding to the comb pilot positions of $\mathbf{D}_{cf4}$, sum the results, combine the peak index value with (\ref{eq21}) to get the velocity estimation $v_2$.
    \item \textbf{Step 4:} We take the average of $r_1$ and $r_2$, and the average of $v_1$ and $v_2$ to get the final estimation.
\end{itemize}

\subsubsection{High and low-frequency full-block aggregated pilot}
The steps of signal processing are as follows.
\begin{itemize}
    \item \textbf{Step 1:} The low-frequency channel information matrix $\mathbf{D}_{cf5}$ and the high-frequency channel information matrix $\mathbf{D}_{cf6}$ are obtained.
    \item \textbf{Step 2:} Since both channel information metrics are block-pilot structures, only the processing of $\mathbf{D}_{cf5}$ is provided here and the processing of $\mathbf{D}_{cf6}$ is consistent. 
    We perform IDFT on the column vectors corresponding to the block pilot positions of $\mathbf{D}_{cf5}$, sum the results, search the power spectrum for the peak index value, and combine it with (\ref{eq19}) to get the range estimation $r_3$. Likewise, we perform CS-DFT on every row vector of $\mathbf{D}_{cf5}$, sum the results, combine the peak index value with (\ref{eq21}) to get the velocity estimation $v_3$. 
    Then, the same process is performed on $\mathbf{D}_{cf6}$ to 
    get $r_4$ and $v_4$.
    \item \textbf{Step 3:} We take the average of $r_3$ and $r_4$, 
    and the average of $v_3$ and $v_4$ to get the final estimation.
\end{itemize}

\subsubsection{High and low-frequency full-comb aggregated pilot}
The steps of signal processing are as follows.
\begin{itemize}
    \item \textbf{Step 1:} The low-frequency channel information matrix $\mathbf{D}_{cf7}$ and the high-frequency channel information matrix $\mathbf{D}_{cf8}$ are obtained.
    \item \textbf{Step 2:} Since both channel information metrics are comb-pilot structures, only the processing of $\mathbf{D}_{cf7}$ is provided here, and the processing of $\mathbf{D}_{cf8}$ is consistent. We perform CS-IDFT on every column vector of $\mathbf{D}_{cf7}$, sum the results, search the power spectrum for the peak index value, and combine it with (\ref{eq19}) to get the range estimation $r_5$. 
    Likewise, we perform DFT on the row vectors corresponding to the comb pilot positions of $\mathbf{D}_{cf7}$, sum the results, combine the peak index value with (\ref{eq21}) to get the velocity estimation $v_5$. 
    Then, the same process is performed on $\mathbf{D}_{cf8}$ to get $r_6$ and $v_6$.
    \item \textbf{Step 3:} We take the average of $r_5$ and $r_6$, and the average of $v_5$ and $v_6$ to get the final estimation.
\end{itemize}

To verify the superiority of the staggered pilot structure chosen 
in this paper, 
the RMSE performance of various pilot structures is compared in Section \ref{sec5}.

\subsection{CRLBs of Various Pilot Structures}

The CRLBs of CA-enabled ISAC signals under different pilot structures are derived. 
It is noted that since the carrier frequencies in high and low-frequency bands are different 
and the estimator of Doppler frequency shift $f_d =2v_0f_c/c_0$ contains the carrier frequency $f_c$. 
To facilitate theoretical derivation, we take out the carrier frequency $f_c$ and 
set the estimator as $ \theta= 2v_0/c_0 $. 
In this way, when performing the derivation, 
we do not have to pay attention to the effect of the different carrier frequencies 
in high and low-frequency bands.
The received echo signal at RX can be expressed as
\begin{equation}\label{eq22}
	y_{m,n} = h_{m,n} e^{j2\pi m T_s f_c \theta} {e^{ -j2\pi n \Delta f \tau}} +\omega_{m,n},
\end{equation}
where $ h_{m,n} $ denotes the product of the channel attenuation factor and 
the amplitude of transmitted signal, 
$ \omega_{m,n} \sim \mathcal{N}\left(0, \sigma^2\right)  $ is the AWGN,
$ m $ represents the index of $ m $-th symbol, 
and $ n $ represents the index of $ n $-th subcarrier. 
It is revealed that $ \tau $ and $ \theta $ are unknown in the received echo signal, 
which contain the range and velocity information of target. 
Therefore, joint estimation of parameters $ \tau $ and $ \theta $ is required. 
The log-likelihood function is 
 \begin{equation}\label{eq23}
	\begin{aligned}
		\ln  p(y;\tau,\theta) =& -\frac{MN}{2}\ln(2\pi \sigma^2) \\
		& -\frac{1}{{2 \sigma}^2} \sum\limits_{m}\sum\limits_{n}(y_{m,n}-s_{m,n})^\ast (y_{m,n}-s_{m,n}),
	\end{aligned}
\end{equation}
where $(\cdot)^{\ast}$ denotes the complex conjugate of the complex number and  
\begin{equation}\label{eq24}
	\begin{aligned}
		p(y;\tau,\theta)=&\frac{1}{(2\pi \sigma^2)^{MN/2}}  \\
		& \cdot e^{-\frac{1}{{2 \sigma}^2} \sum\limits_{m}\sum\limits_{n} \left|  y_{m,n} - h_{m,n} e^{j2\pi m T_s f_c \theta} {e^{ -j2\pi n \Delta f \tau}} \right|^2 },
	\end{aligned}
\end{equation}
\begin{equation}\label{eq25}
	s_{m,n} =  h_{m,n} e^{j2\pi m T_s f_c \theta} {e^{ -j2\pi n \Delta f \tau}}.
\end{equation}

The first derivative of $ \tau $ and $ \theta $ can be expressed as
{\small \begin{equation}\label{eq26}
	\begin{aligned}
		&\frac{\partial \ln p(y;\tau,\theta)}{\partial \tau} = \frac{-1}{{2 \sigma}^2} \sum\limits_{m}\sum\limits_{n}  \\
		&\cdot \left[(y_{m,n}-s_{m,n})^\ast(s_{m,n})-(y_{m,n}-s_{m,n})(s_{m,n})^\ast \right](j2\pi n \Delta f),
	\end{aligned}
\end{equation}
\begin{equation}\label{eq27}
	\begin{aligned}
		&\frac{\partial \ln p(y;\tau,\theta)}{\partial \theta} = \frac{1}{{2 \sigma}^2} \sum\limits_{m}\sum\limits_{n}  \\
		&\cdot \left[(y_{m,n}-s_{m,n})^\ast(s_{m,n})-(y_{m,n}-s_{m,n})(s_{m,n})^\ast \right](j2\pi mT_sf_c).
	\end{aligned}
\end{equation}}

The second derivative of $ \tau $ is
{\small \begin{equation}\label{eq28}
	\begin{aligned}
		\frac{\partial^2 \ln p(y;\tau,\theta)}{\partial \tau^2} =& -\frac{1}{{2 \sigma}^2} \sum\limits_{m}\sum\limits_{n}(j2\pi n\Delta f) \\
		& \cdot \left [  \left ( y_{m,n}^{\ast }-s_{m,n}^{\ast } \right )\frac{\partial s_{m,n}}{\partial \tau } -\left ( s_{m,n} \right )\frac{\partial s_{m,n}^{\ast }}{\partial \tau }  \right. \\
		&{-} \left. \left ( y_{m,n}-s_{m,n} \right )\frac{\partial s_{m,n}^{\ast }}{\partial \tau } +\left ( s_{m,n}^{\ast } \right )\frac{\partial s_{m,n}}{\partial \tau }   \right ].
	\end{aligned}
\end{equation}}

Upon taking the negative expectation of (\ref{eq28}), we have
{\small \begin{equation}\label{eq29}
	\begin{aligned}
		&{F}_{1,1} = -E\left( \frac{\partial^2 \ln p(y;\tau,\theta)}{\partial \tau^2}\right) \\
		&= \frac{1}{{2 \sigma}^2} \sum\limits_{m}\sum\limits_{n}(j2\pi n\Delta f)  
		\left [ \left ( -s_{m,n} \right )\frac{\partial s_{m,n}^{\ast }}{\partial \tau } + \left ( s_{m,n}^{\ast} \right )\frac{\partial s_{m,n} }{\partial \tau } \right ] \\
		&= \frac{1}{\sigma^2} \sum\limits_{m}\sum\limits_{n}(2\pi n \Delta f)^2 h_{m,n}^{2}. 
	\end{aligned}
\end{equation}}

Similarly, 
the negative expectation of the second derivative of $ \theta $ can be obtained
{\small \begin{equation}\label{eq30}
	\begin{aligned}
		{F}_{2,2} &= -E\left( \frac{\partial^2 \ln p(y;\tau,\theta)}{\partial \theta^{2}}\right) \\
		&= \frac{1}{\sigma^2} \sum\limits_{m}\sum\limits_{n}(2\pi m T_s f_c)^2 h_{m,n}^{2},
	\end{aligned}
\end{equation}}
{\small \begin{equation}\label{eq31}
	\begin{aligned}
		{F}_{1,2}=F_{2,1} &= -E\left(  \frac{\partial^2 \ln p(y;\tau,\theta)}{\partial \tau \partial \theta}\right) \\
		&= \frac{-1}{\sigma^2} \sum\limits_{m}\sum\limits_{n}(2\pi)^2 h_{m,n}^{2} m n \Delta f T_s f_c. 
	\end{aligned}
\end{equation}}

Then, the Fisher information matrix can be derived as
{\small \begin{equation}\label{eq32}
	\begin{aligned}
		\textbf{F}^{-1} &= \left[
		\begin{array}{cc}
			CRLB(\tau)	& CRLB(\tau,\theta)  \\
			CRLB(\theta,\tau)	& CRLB(\theta)
		\end{array}
		\right]\\
		&=\left[
		\begin{array}{cc}
			F_{1,1}	& F_{1,2} \\
			F_{2,1}	& F_{2,2}
		\end{array}
		\right]^{-1},
	\end{aligned}
\end{equation}}
and
{\small \begin{equation}\label{eq33}
	\begin{aligned}
		& CRLB(\tau)  = \frac{F_{2,2}}{F_{1,1}F_{2,2}-F_{1,2}F_{2,1}} \\
		& = \frac{\sigma^2}{(2\pi)^2 h_{m,n}^{2}} \cdot 
		\frac{1}{\sum\limits_{m}\sum\limits_{n}(n\Delta f)^2 - \frac{\left(\sum\limits_{m}\sum\limits_{n}mn\Delta f T_s f_c\right)^2}{\sum\limits_{m}\sum\limits_{n}\left(m T_s f_c\right)^2} },
	\end{aligned}
\end{equation}
\begin{equation}\label{eq34}
	\begin{aligned}
		& CRLB(\theta) = \frac{F_{1,1}}{F_{1,1}F_{2,2}-F_{1,2}F_{2,1}} \\ 
		& = \frac{\sigma^2}{(2\pi)^2 h_{m,n}^{2}} \cdot
		\frac{1}{\sum\limits_{m}\sum\limits_{n}(m T_s f_c)^2 - \frac{\left(\sum\limits_{m}\sum\limits_{n}mn\Delta f T_s f_c\right)^2}{\sum\limits_{m}\sum\limits_{n}\left(n \Delta f\right)^2} }.
	\end{aligned}
\end{equation}}

According to the transformation relation of $ \tau= 2R/c_0  $ and $ \theta = 2v/c_0 $, 
the CRLBs of range and velocity estimation are  
\begin{equation}\label{eq35}
	CRLB(R)=\frac{c_{0}^{2}}{4}CRLB(\tau),
\end{equation}
\begin{equation}\label{eq36}
	CRLB(v)=\frac{c_{0}^{2}}{4}CRLB(\theta).
\end{equation}

\subsubsection{CRLB of CA-based staggered pilot signal}
According to Section \ref{s3}, 
the interval of comb pilot satisfies $ K=\Delta f_2/\Delta f_1 $, 
so that we have $ \Delta f_1 K=\Delta f_2 $ in the derivation of CRLB.
According to \eqref{eq33} and \eqref{eq35}, 
the CRLB of range estimation $ CRLB(R) $ is 
{\small  \begin{equation}\label{eq37}
	\begin{aligned}
		CRLB(R) =&\frac{c_0^{2}}{4} \\
		&  \cdot \frac{\sigma^2}{(2\pi)^2 h_{m,n}^{2}} \cdot 
		\frac{1}{\sum\limits_{m}\sum\limits_{n}(n\Delta f)^2 - \frac{\left(\sum\limits_{m}\sum\limits_{n}mn\Delta f T_s\right)^2}{\sum\limits_{m}\sum\limits_{n}\left(m T_s\right)^2} }.
	\end{aligned}
\end{equation}}

Then, combining with the process of Algorithm 1, 
the sum in \eqref{eq37} is the sum of two channel information matrices in high and 
low-frequency bands. 
Therefore,
\begin{equation}\label{eq38}
	\begin{aligned}
		&\sum\limits_{m=0}^{M-1}\sum\limits_{n=0}^{N_\alpha-1}(nK \Delta f_1)^2_{\mathrm{Low}}+\sum\limits_{m=0}^{M_\beta-1}\sum\limits_{n=0}^{N-1}(n \Delta f_2)^2_{\mathrm{High}} =  \\ & \tfrac{MN_{\alpha}(N_{\alpha}-1)(2N_{\alpha}-1){\Delta f}_{1}^{2}K^2}{6} + 
		\tfrac{M_{\beta}N(N-1)(2N-1){\Delta f}_{2}^{2}}{6},
	\end{aligned}
\end{equation}
\begin{equation}\label{eq39}
	\begin{aligned}
		\scriptscriptstyle&\sum\limits_{m=0}^{M-1}\sum\limits_{n=0}^{N_\alpha-1}(mT_1f_{c1})^2_{\mathrm{Low}}+\sum\limits_{m=0}^{M_\beta-1}\sum\limits_{n=0}^{N-1}(mQT_2f_{c2})^2_{\mathrm{High}} =\\& \tfrac{N_{\alpha}M(M-1)(2M-1)T_{1}^{2}f_{c1}^{2}}{6} +
		\tfrac{NQ^2 M_{\beta}(M_{\beta}-1)(2M_{\beta}-1)T_{2}^{2}f_{c2}^{2}}{6},
	\end{aligned}
\end{equation}
{\small \begin{equation}\label{eq40} 
		\begin{aligned}
			 & \sum\limits_{m=0}^{M-1}\sum\limits_{n=0}^{N_\alpha-1}(mnK\Delta f_1 T_1 f_{c1})_{\mathrm{Low}}+\sum\limits_{m=0}^{M_\beta-1}\sum\limits_{n=0}^{N-1}(mQn\Delta f_2 T_2 f_{c2})_{\mathrm{High}} \\ & = \tfrac{\Delta f_{1}KT_1f_{c1} N_{\alpha}(N_{\alpha}-1)M(M-1)}{4} +
			\tfrac{\Delta f_{2}T_2f_{c2} N(N-1)M_{\beta}(M_{\beta}-1)Q}{4},
	\end{aligned}
\end{equation}}
where $ N_{\alpha}=N/K $ is the number of subcarriers occupied by the pilot in low-frequency band, 
$ M_{\beta}=M/Q $ is the number of OFDM symbols occupied by the pilot in high-frequency band. 
By substituting \eqref{eq38}, \eqref{eq39}, and \eqref{eq40} into \eqref{eq37}, 
the CRLB of range estimation of CA-based staggered pilot signal is \eqref{eq41}.
\begin{figure*}[!ht]
	\begin{align}
		 CRLB(R)_1= \tfrac{\tfrac{3c_0^{2}\sigma^2}{8\pi^2 \Delta f_{2}^{2} h_{m,n}^{2}}}{\left[ MN_{\alpha}(N_{\alpha}-1)(2N_{\alpha}-1) + 
				M_{\beta}N(N-1)(2N-1) -\tfrac{9\left[  N_{\alpha}(N_{\alpha}-1)M(M-1) + 
					N(N-1)M_{\beta}(M_{\beta}-1)Q \right]^2 }{4\left[N_{\alpha}M(M-1)(2M-1) + 
					NQ^2 M_{\beta}(M_{\beta}-1)(2M_{\beta}-1)\right]} \right]}.
		\label{eq41}
	\end{align}
    \begin{align}
		CRLB(v)_1= \tfrac{\tfrac{3c_0^{2}\sigma^2}{8\pi^2 f_{c1}^2 T_{1}^2 h_{m,n}^{2}}}{\left[N_{\alpha}M(M-1)(2M-1) + NQ^2 M_{\beta}(M_{\beta}-1)(2M_{\beta}-1) -\tfrac{9\left[ N_{\alpha}(N_{\alpha}-1)M(M-1) + 
				N(N-1)M_{\beta}(M_{\beta}-1)Q \right]^2 }{4\left[ MN_{\alpha}(N_{\alpha}-1)(2N_{\alpha}-1) + 
				M_{\beta}N(N-1)(2N-1) \right]} \right]}.
		\label{eq43}
	\end{align}
		{\noindent} \rule[-10pt]{18cm}{0.05em}
\end{figure*}

In terms of velocity estimation, 
$T_1f_{c1}=T_2f_{c2}$ is known using Algorithm 2, 
so that only one pair of parameters, namely $ T_1 $ and $ f_{c1} $, are selected when deriving CRLB.
Firstly, (\ref{eq42}) can be obtained 
according to \eqref{eq34} and \eqref{eq36}.

{\small \begin{equation}\label{eq42}
	\begin{aligned}
		CRLB(v) =&\frac{c_0^{2}}{4}  \\
		& \cdot \frac{\sigma^2}{(2\pi)^2 h_{m,n}^{2}} \cdot
		\frac{1}{\sum\limits_{m}\sum\limits_{n}(m T_sf_c)^2 - \frac{\left(\sum\limits_{m}\sum\limits_{n}mn\Delta f T_sf_c\right)^2}{\sum\limits_{m}\sum\limits_{n}\left(n \Delta f\right)^2} }.
	\end{aligned}
\end{equation}}

Then, according to Algorithm 2, 
the sum in \eqref{eq42} is the sum of two channel information matrices in high and 
low-frequency bands. 
Similarly, substituting \eqref{eq38}, \eqref{eq39}, and \eqref{eq40} into \eqref{eq42}, 
the CRLB of velocity estimation of the CA-based staggered pilot ISAC signal 
is shown in \eqref{eq43}.
\begin{figure*}[!htb]
	\begin{align}
		  CRLB(R)_2= \tfrac{\tfrac{3c_0^{2}\sigma^2}{8\pi^2  h_{m,n}^{2}}}{\left[ \Delta f_{2}^{2} K^2 MN_{\alpha}(N_{\alpha}-1)(2N_{\alpha}-1) + 
				\Delta f_{1}^{2}	M_{\beta}N(N-1)(2N-1) -\tfrac{9\left[ \Delta f _2 K N_{\alpha}(N_{\alpha}-1)M(M-1) + 
					\Delta f_1 N(N-1)M_{\beta}(M_{\beta}-1)Q \right]^2 }{4\left[N_{\alpha}M(M-1)(2M-1)+ 
					NQ^2 M_{\beta}(M_{\beta}-1)(2M_{\beta}-1)\right]} \right]} .
		\label{eq44}
	\end{align}
	\begin{align}
		  CRLB(v)_2=  \tfrac{\tfrac{3c_0^{2}\sigma^2}{8\pi^2 f_{c1}^2 T_{1}^2 h_{m,n}^{2}}}{\left[N_{\alpha}M(M-1)(2M-1) + NQ^2 M_{\beta}(M_{\beta}-1)(2M_{\beta}-1) -\tfrac{9\left[ \Delta f_{2} K N_{\alpha}(N_{\alpha}-1)M(M-1) + 
					\Delta f_{1} N(N-1)M_{\beta}(M_{\beta}-1)Q \right]^2 }{4\left[ \Delta f _{2}^{2} K^2 MN_{\alpha}(N_{\alpha}-1)(2N_{\alpha}-1) + 
					\Delta f_{1}^{2} M_{\beta}N(N-1)(2N-1) \right]} \right]} .
		\label{eq45}
	\end{align}
\end{figure*}

The CRLBs for the CA-enabled ISAC signals under the remaining pilot structures are 
provided as follows. 
For the simplicity in deriving the CRLBs, 
the comb pilot intervals of low and high-frequency bands are set the same, which is
denoted by $ K $.
Meanwhile, the block pilot intervals of low and high-frequency bands are set the same, 
denoting by $ Q $. 

\subsubsection{CRLB with high-frequency comb and low-frequency block aggregated pilot}
Similarly, the CRLBs for range and velocity estimation are 
given by \eqref{eq44} and \eqref{eq45}, respectively. 

\subsubsection{CRLB with high and low-frequency full-block aggregated pilot}
The CRLBs for range and velocity estimation are 
given by \eqref{eq46} and \eqref{eq47}, respectively.

\subsubsection{CRLB with high and low-frequency full-comb aggregated pilot}
The CRLBs for range and velocity estimation are 
given by \eqref{eq48} and \eqref{eq49}, respectively.

\begin{figure*}[!htb]
	\begin{align} 
		{  CRLB(R)_3= \tfrac{\tfrac{3c_0^{2}\sigma^2}{8\pi^2  h_{m,n}^{2}} }{N(N-1)M_{\beta}\left[ (\Delta f_{1}^{2}+\Delta f_{2}^{2})(2N-1) - \tfrac{ 9\left(\Delta f_1+\Delta f_2 \right)^2(N-1) (M_{\beta}-1) }{8(2M_{\beta}-1)} \right]}}.
		\label{eq46}
	\end{align}
	\begin{align}
		{ CRLB(v)_3= \tfrac{\tfrac{3c_0^{2}\sigma^2}{8\pi^2 f_{c1}^2 T_{1}^2 h_{m,n}^{2}}}{NM_{\beta}(M_{\beta}-1)\left[2(2M_{\beta}-1)-\tfrac{9\left(\Delta f_1+\Delta f_2 \right)^2(N-1)(M_{\beta}-1) }{4\left( \Delta f_1^{2}+\Delta f_2^{2}\right)(2N-1) }\right]}}.
		\label{eq47}
	\end{align}
	\begin{align}
		CRLB(R)_4= \tfrac{\tfrac{3c_0^{2}\sigma^2}{8\pi^2 \Delta f_2^{2} h_{m,n}^{2}}}{N_{\alpha}(N_{\alpha}-1)M\left[ (2N_{\alpha}-1)(K^2+1) - \tfrac{9\left( 1 +K \right)^2 (N_{\alpha}-1) (M-1) }{8 (2M-1)} \right]}.
		\label{eq48}
	\end{align}
	\begin{align}
		CRLB(v)_4= \tfrac{\tfrac{3c_0^{2}\sigma^2}{8\pi^2 f_{c1}^2 T_{1}^2 h_{m,n}^{2}} }{N_{\alpha}(M-1)M\left[ 2(2M-1) - \tfrac{9\left( 1 +K  \right)^2 (N_{\alpha}-1) (M-1) }{4\left(1+K^{2} \right) (2N_{\alpha}-1)} \right]}.
		\label{eq49}
	\end{align}
	{\noindent} \rule[-10pt]{18cm}{0.05em}
\end{figure*}

\section{Simulation Results and Analysis}\label{sec5}

\begin{table*}
	\caption{Parameter Configuration of ISAC System based on CA \cite{knill2018high,braun2009parametrization,3gpp2018nr}.}
	\label{tab_3}
	\renewcommand{\arraystretch}{1.55} 
	\begin{center}
		\begin{tabular}{|m{0.1\textwidth}<{\centering}| m{0.3\textwidth}<{\centering}| m{0.15\textwidth}<{\centering}| m{0.15\textwidth}<{\centering}|}
			\hline
			\textbf{Symbol} & \textbf{Parameter} & \textbf{Low frequency} & \textbf{High frequency} \\
			\hline
			$ f_c $	& Carrier frequency &  5.9\;GHz  &  24\;GHz  \\
			\hline
			$ M $	& Number of OFDM symbols &  64 &  64 \\
			\hline
			$ N $	& Number of subcarriers &  512  & 512  \\
			\hline
			$ K $ & Comb pilot interval &  4  &  4  \\
			\hline
			$ Q $	& Block pilot interval &  4  &   4 \\
			\hline
			$ \Delta f $	& Subcarrier spacing &  30\;kHz  & 120\;kHz \\
			\hline
			$ T $	& Total OFDM symbol duration  &  39.5\;$\mathrm{\mu}$s   & 9.7\;$\mathrm{\mu}$s  \\
			\hline
			$ B $	& Bandwidth of signal  &  15.4\;MHz  &  61.4\;MHz  \\
			\hline

		\end{tabular}
	\end{center}
\end{table*}

In this section, 
the power spectrum and RMSE of radar sensing are  
simulated to verify the feasibility and performance of the proposed CA-based
ISAC signal design and processing methods. 
Then, the RMSEs and CRLBs of the four pilot structures are simulated.

Firstly, we assume that the maximum unambiguous velocity of sensing is $v=50$ m/s. 
The subcarrier spacing in the 5.9 GHz low-frequency band is 30 kHz, 
while the subcarrier spacing in the 24 GHz high-frequency band is 120 kHz according to the 3GPP TS 38.211 standard \cite{3gpp2018nr} and subcarrier spacing design principle in \cite{braun2009parametrization}.
With the farthest detection distance 200 m, 
the length of CP must be greater than the maximum multipath delay spread, 
so that the length of CP must be greater than 1.33 $\mu$s and there is no inter-symbol interference (ISI). 
The parameters in simulation are summarized in Table \ref{tab_3}, 
which satisfies the scenario of vehicle communication and sensing.

\subsection{Performance of Radar Sensing}

\begin{figure}[!htbp]
    \centering
	\includegraphics[width=0.42\textwidth]{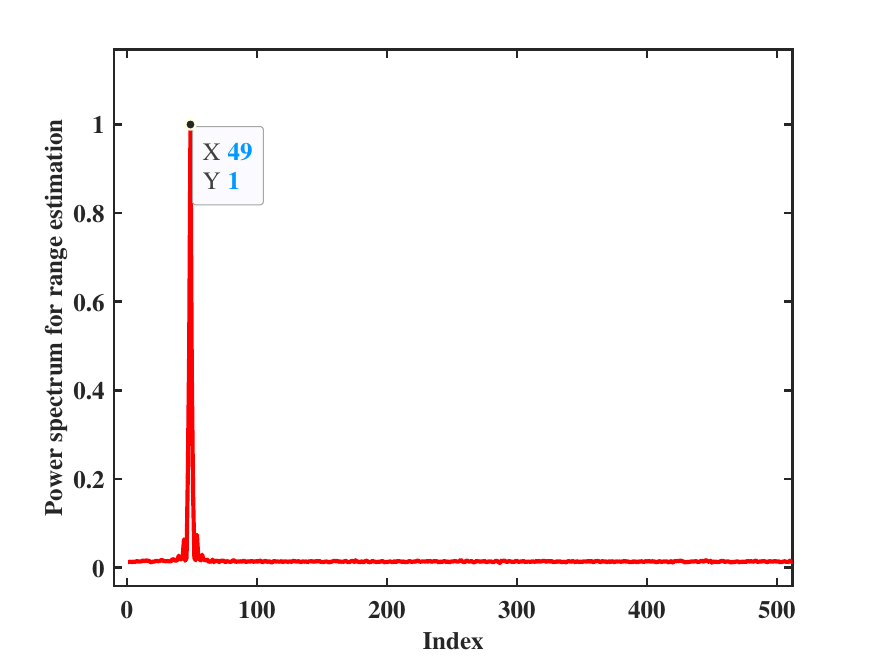}
	\caption{Power spectrum of range.}
	\label{range power}
\end{figure}

\begin{figure}[!htbp]
	\centering
	\includegraphics[width=0.42\textwidth]{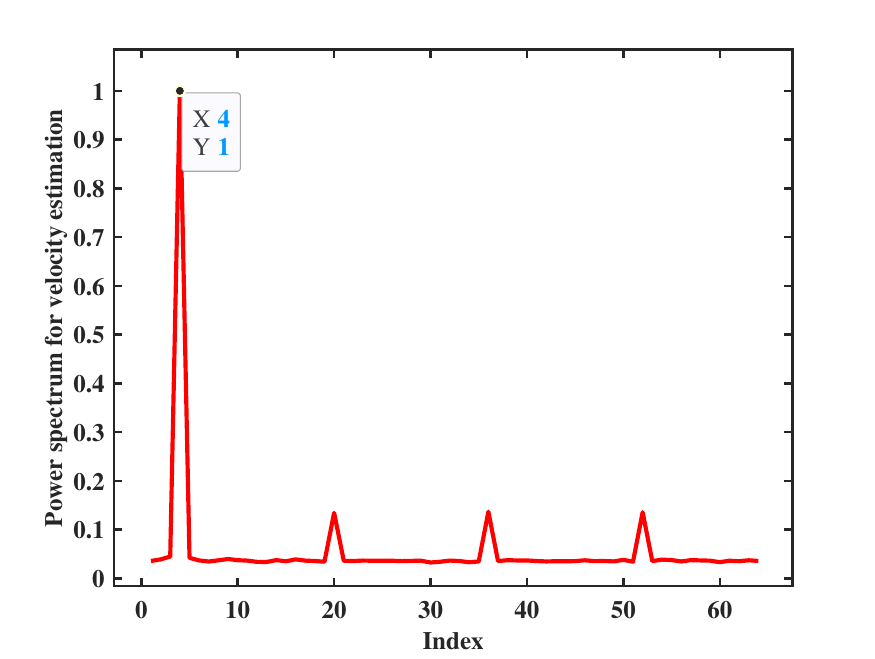}
	\caption{Power spectrum of velocity.}
	\label{velocity power}
\end{figure}

In this subsection, 
the feasibility and sensing performance of the proposed ISAC signal processing algorithm 
are verified by simulating the power spectrum and RMSE of radar sensing 
according to the parameters in Table \ref{tab_3}. 
The power spectra for range and velocity estimation are simulated
with the range of target 117 m, the relative velocity of target 30 m/s and 
the SNR 10 dB, 
as shown in Fig. \ref{range power} and Fig. \ref{velocity power}, respectively. 
$\text{X}$ and $\text{Y}$ in the figure are used to indicate the peak and the corresponding index value. As the power spectrum is normalized, so that it corresponds to the peak position when $\text{Y}=1$.
According to the indexes of the peaks of the power spectra, 
the target's range and velocity are estimated as follows.

\begin{equation}\label{eq50}
	\widehat{R}= \dfrac{(ind_n -1)c_0}{2\Delta f N } = 117.1875\;\rm{m},
\end{equation}

\begin{equation}\label{eq51}
	\widehat{v_0}=\dfrac{(ind_m-1)c_0}{2f_{c2}T_2M} = 30.3176\;\rm{m/s},
\end{equation}
where $ind_n=49$ and $ind_m=4$ represent the peak index value of 
the power spectra of range and velocity, respectively.

It is revealed that the error of range estimation is 0.1875 m
and the error of velocity estimation is 0.3176 m/s.
Then, RMSE is used to reveal the deviation between 
the estimated value and the real value. 
The performance of range and velocity estimation using high-frequency signal 
is better than that using low-frequency signal. 
Meanwhile, the block pilot signal performs better 
in range estimation and the comb pilot siganl performs better in velocity estimation.  
Hence, the CA-based staggered pilot signal is compared with 
high-frequency block pilot signal in range estimation 
and is compared with high-frequency comb pilot signal 
in velocity estimation to verify the performance improvement using CA. 

\begin{figure}[!htbp]
	\centering
	\includegraphics[width=0.43\textwidth]{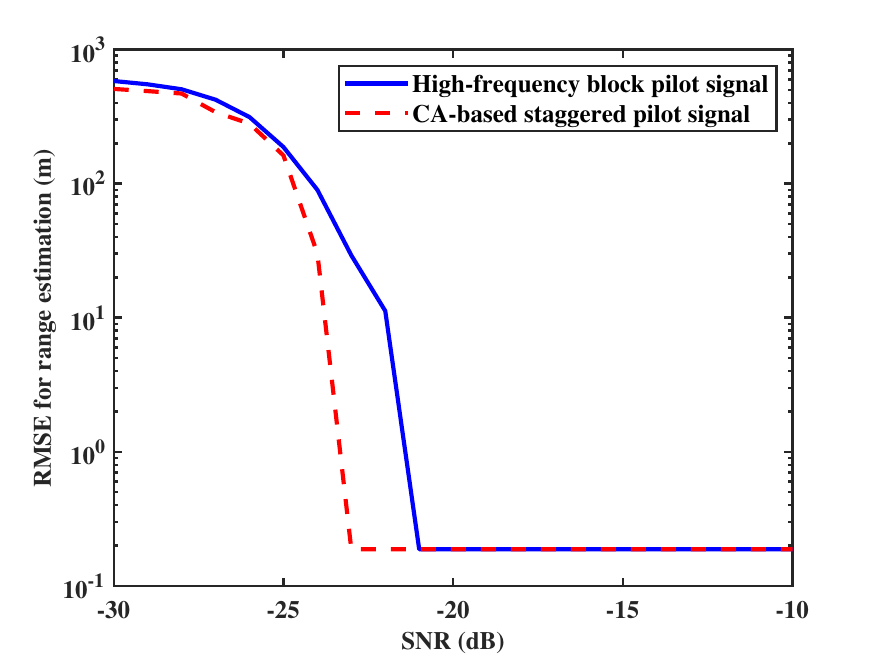}
	\caption{RMSE for the range.}
	\label{range rmse}
\end{figure}
\begin{figure}[!htbp]
	\centering
	\includegraphics[width=0.43\textwidth]{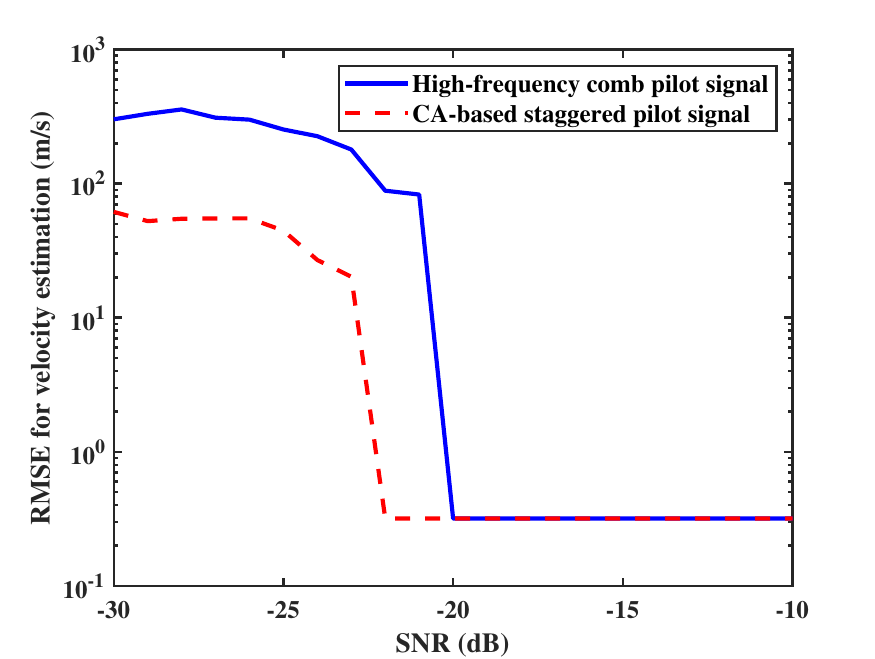}
	\caption{RMSE for the velocity.}
	\label{velocity rmse}
\end{figure}

The RMSE of the high-frequency signal and the CA-based staggered pilot signal 
for range and velocity estimation are shown 
in Fig. \ref{range rmse} and Fig. \ref{velocity rmse}, respectively.
It is revealed that 
the RMSEs of the range and velocity estimation with the proposed ISAC signal, 
namely the CA-based staggered pilot signal,
are smaller than those with high-frequency block pilot signal and high-frequency comb pilot signal, respectively.
Hence, the CA-based staggered pilot signal has better 
anti-noise capability and sensing performance than the signals without CA.

\subsection{CRLB of Range and Velocity Estimation}

The CRLBs of CA-based staggered pilot ISAC signals is simulated.
Then, the comparison of Root CRLBs (abbreviated as RCRLBs) and RMSE is simulated 
based on the parameters in Table \ref{tab_3}.

\subsubsection{CRLB analysis}

\begin{figure}[!htbp]
	\centering
	\includegraphics[width=0.43\textwidth]{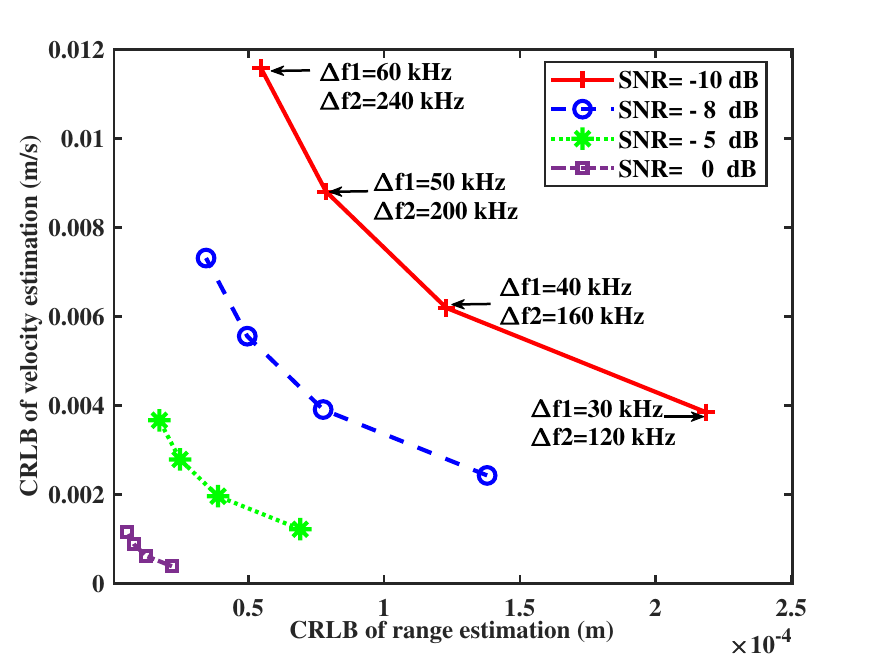}
	\caption{CRLB for range and velocity with different subcarrier spacing and different SNR.}
	\label{CRLB analysis}
\end{figure}

As shown in Fig. \ref{CRLB analysis}, 
when the SNR is fixed, 
the CRLB of range estimation is decreasing with the increase of 
the CRLB of velocity estimation, namely, the CRLBs of range and velocity estimation have a tradeoff relation. 
The reason is that the CRLB of velocity estimation is proportional to the subcarrier spacing, 
whereas the CRLB of range estimation is inversely proportional to the subcarrier spacing. 
Besides, when the subcarrier spacing is fixed, 
the CRLBs of range and velocity estimation 
are decreasing with the increase of SNR.

\subsubsection{CRLB comparison}

\begin{figure}[!htbp]
	\centering
	\includegraphics[width=0.43\textwidth]{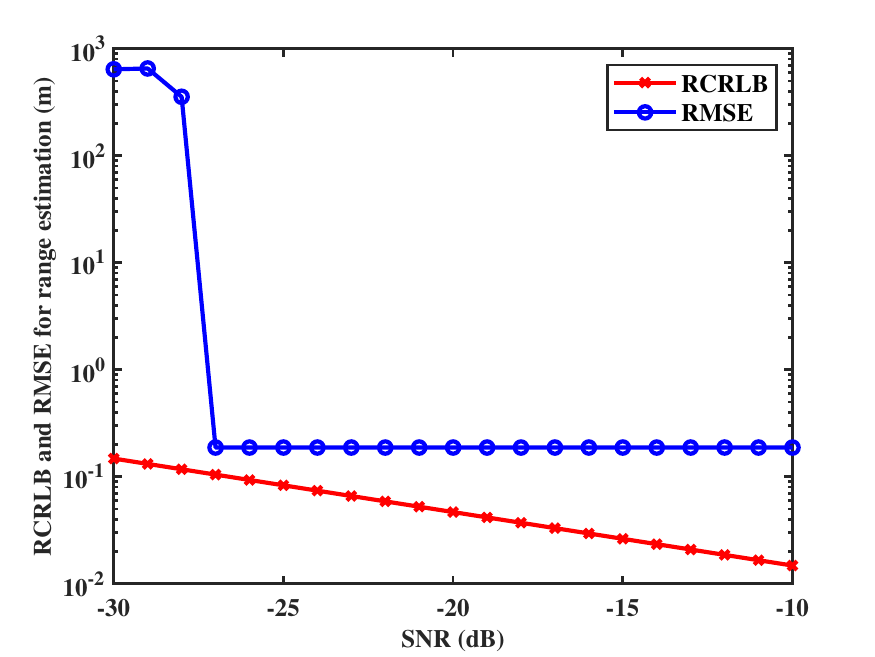}
	\caption{Comparison of RCRLB and RMSE for range.}
	\label{mse range}
\end{figure}

\begin{figure}[!htbp]
	\centering
	\includegraphics[width=0.43\textwidth]{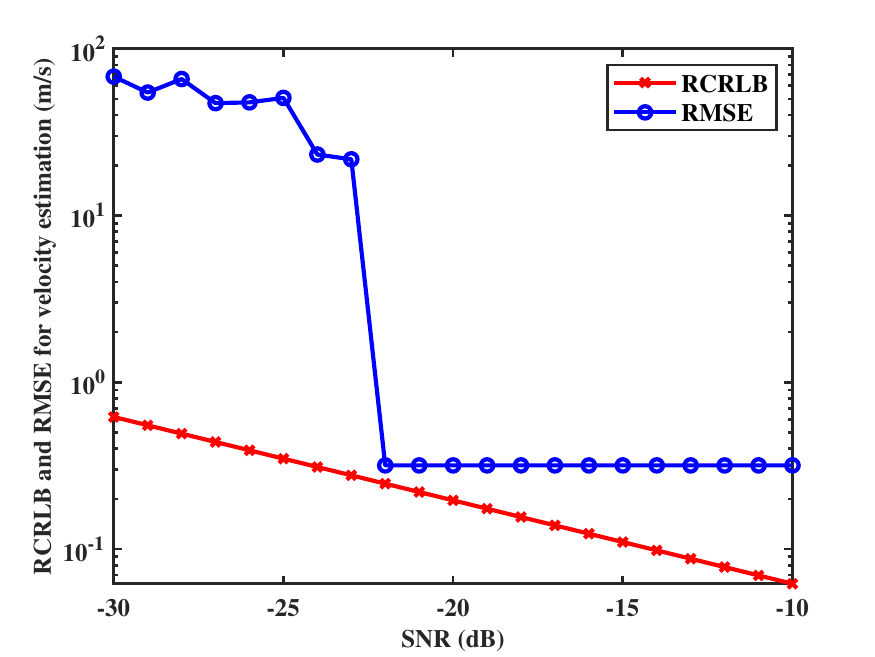}
	\caption{Comparison of RCRLB and RMSE for velocity.}
	\label{mse velocity}
\end{figure}

Fig. \ref{mse range} and Fig. \ref{mse velocity} show the comparison of 
RCRLB and RMSE for range and velocity estimation, respectively. 
It is observed that the RCRLB is smaller than the RMSE 
because the RCRLB reveals the lower bound of the minimum root variance of unbiased estimation. 
Meanwhile, the 2D-FFT algorithm has a fixed resolution 
due to the fixed sampling rate and number of FFT points in the simulation, 
causing the RMSE to converge to a fixed value in the high SNR regime.


\subsection{Simulation Comparison of Four Pilot Structures}
In this subsection, we simulate the RMSEs and CRLBs for four pilot structures based on the parameters in Table \ref{tab_3}.
\subsubsection{CRLBs of four pilot structures}
\begin{figure}[!htbp]
 	\centering
 	\includegraphics[width=0.42\textwidth]{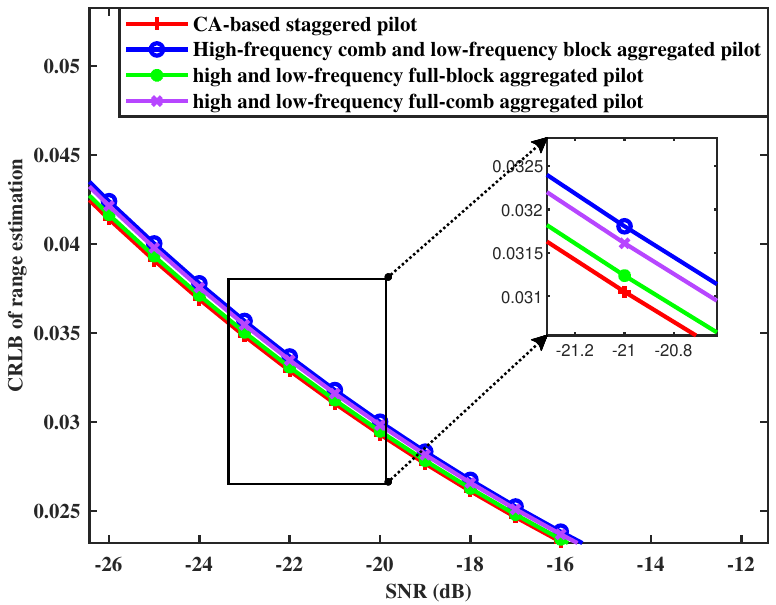}
 	\caption{Range CRLB simulation comparison diagram of four aggregation types.}
 	\label{CRLB r}
 \end{figure}

\begin{figure}[!htbp]
 	\centering
 	\includegraphics[width=0.42\textwidth]{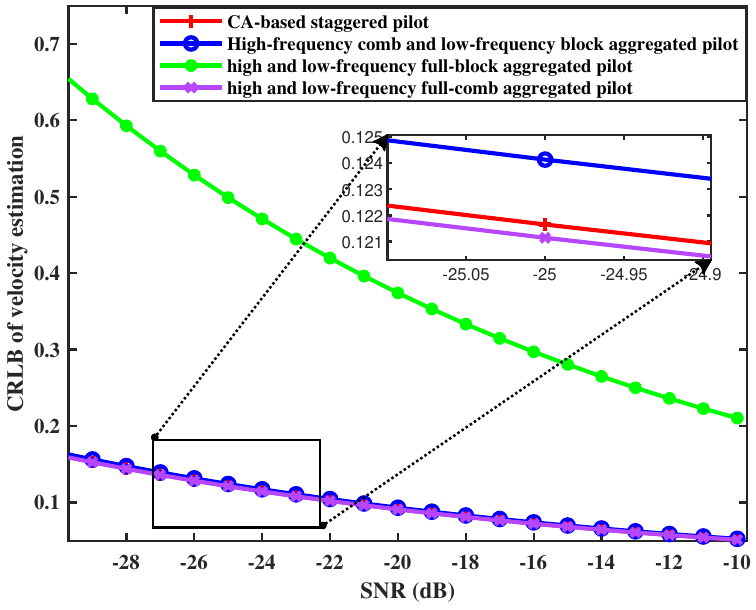}
 	\caption{Velocity CRLB simulation comparison diagram of four aggregation types.}
 	\label{CRLB v}
 \end{figure}

According to (\ref{eq41}) - (\ref{eq49}), the CRLBs for range and velocity estimation with the four pilot structures 
are shown in Fig. \ref{CRLB r} 
and Fig. \ref{CRLB v}, respectively. 
As revealed in Fig. \ref{CRLB r}, 
the CA-based staggered pilot has the lowest CRLB. 
Therefore, when the ISAC-enabled mobile communication system is mainly used to 
obtain high-accurate range estimation of target, 
the CA-based staggered pilot structure is the optimal choice. 
In terms of velocity estimation, as shown in Fig. \ref{CRLB v}, 
the high and low-frequency full-comb aggregated pilot has the lowest CRLB, 
followed by the CA-based staggered pilot. 
It is observed that the gap of the CRLBs between the low-frequency full-comb aggregated pilot and the CA-based staggered pilot is small. 
Therefore, when the ISAC-enabled mobile communication system is 
mainly used to obtain high-accurate velocity estimation of target, 
both the high and low-frequency full-comb aggregated pilot and 
the CA-based staggered pilot structure can be chosen.

CRLB comparisons are performed theoretically. 
In practice, the sensing performance of various pilot structures 
depends on the signal processing algorithms. 
Therefore, in the following, 
we simulate the RMSE of four pilot structures using the algorithm proposed in this paper.


\subsubsection{RMSEs of four pilot structures}
\begin{figure}[!htbp]
 	\centering
 	\includegraphics[width=0.45\textwidth]{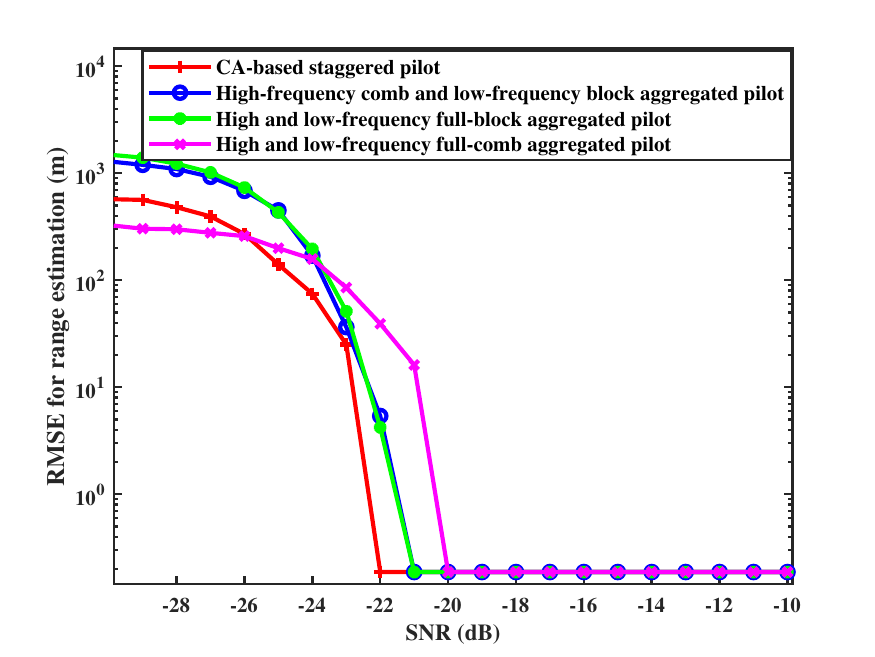}
 	\caption{Range RMSE simulation comparison diagram of four aggregation types.}
 	\label{rmse r}
 \end{figure}

\begin{figure}[!htbp]
 	\centering
 	\includegraphics[width=0.45\textwidth]{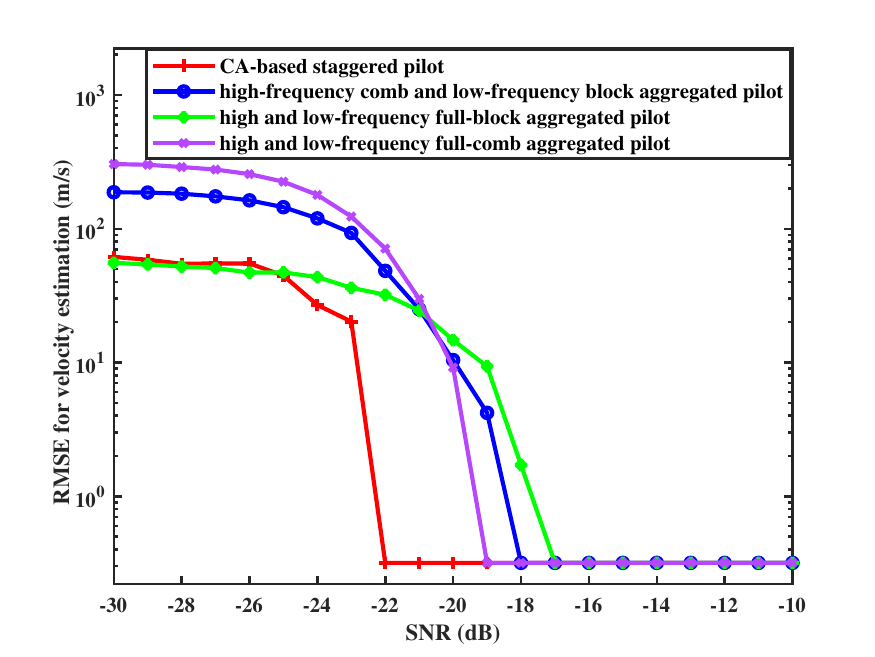}
 	\caption{Velocity RMSE simulation comparison diagram of four aggregation types.}
 	\label{rmse v}
 \end{figure}

The RMSEs for range and velocity estimation with the four pilot structures 
are shown in Fig. \ref{rmse r} 
and Fig. \ref{rmse v}, respectively. 
As revealed in Fig. \ref{rmse r}, 
the CA-based staggered pilot has the fastest convergence speed among 
the four pilot structures. When the SNR is large than the threshold -22 dB, 
the RMSE of range estimation with the CA-based staggered pilot remains constant. 
While the thresholds of SNR with other pilot structures are smaller than -22 dB. 
Hence, in terms of range estimation, 
the CA-based staggered pilot has the best performance among the four pilot structures.

In terms of velocity estimation, as shown in Fig. \ref{rmse v}, 
the CA-based staggered pilot has the fastest convergence speed among the four pilot structures. 
When the SNR is larger than the threshold -22 dB, the RMSE of velocity estimation with 
CA-based staggered pilot is converged to a constant value. 
While the thresholds of SNR with other
pilot structures are smaller than -22 dB. 
Therefore, the CA-based staggered pilot has the best velocity estimation performance among 
the four pilot structures.


\section{Conclusion}

This paper proposes the CA-based ISAC signals 
to improve sensing performance by aggregating discontinuous high and low-frequency bands. 
Then, the radar signal processing algorithms based on 2D-FFT and CS are proposed, 
which 
fuse the sensing information from high and low-frequency bands.
The CRLBs of range and velocity estimation with the proposed ISAC signals are analyzed.
The feasibility and performance improvement of the ISAC signals proposed in this paper 
are verified by simulation results. 
It is revealed that the CA-based ISAC signal has better sensing performance 
compared with the ISAC signal without CA.
Since the full spectrum is one of the promising key technologies in 6G,
which efficiently aggregates the low and high-frequency bands to provide high communication
data rate, the CA-enabled ISAC signals proposed in this 
paper may provide a guideline for the ISAC system design in the era of 6G. 
The future research problems focus on the sensing algorithms with different path losses on high and low frequency bands and the effect of nonlinear high-frequency power amplifier of transmitter RF.


\bibliographystyle{IEEEtran}
\bibliography{CA_reference}
\ifCLASSOPTIONcaptionsoff
\newpage
\fi

\end{document}